\documentclass[apj]{emulateapj}
\usepackage{lineno}
\usepackage{amsmath,amstext}
\usepackage{graphicx}
\usepackage[T1]{fontenc}
\usepackage{apjfonts}
\usepackage[dvipsnames]{xcolor}
\usepackage{color}
\DeclareGraphicsExtensions{.eps,.png,.jpg}
\usepackage{epsfig, psfig}
\usepackage[normalem]{ulem}
\usepackage{threeparttable}
\usepackage[backref,breaklinks,colorlinks,citecolor=blue]{hyperref}
\usepackage[figure,figure*]{hypcap}

\def\asec{$^{\prime\prime}$}

\def\kms{km s$^{-1}$}

\def\sb{mag~arcsec$^{-2}$}

\begin{document}

\title{The MASSIVE SURVEY XVI.  The Stellar Initial Mass Function 
in the Center of MASSIVE Early-Type Galaxies}
\author{
  Meng Gu\altaffilmark{1}, 
  Jenny Greene\altaffilmark{1},
  Andrew B. Newman\altaffilmark{2},
  Christina Kreisch\altaffilmark{1},
  Matthew Quenneville\altaffilmark{3},
  Chung-Pei Ma\altaffilmark{3} 
  and John P. Blakeslee\altaffilmark{4}}  

\altaffiltext{1}{Department of Astrophysical Sciences, Princeton University, 
Princeton, NJ, USA}
\altaffiltext{2}{The Observatories of the Carnegie Institution for Science, 
Pasadena, CA, USA}
\altaffiltext{3}{Department of Astronomy and Department of Physics, University of California at Berkeley, Berkeley, CA, USA}
\altaffiltext{4}{Gemini Observatory and NSF\textquotesingle s NOIRLab, 950 N. Cherry Avenue, Tucson, AZ 85719, USA}
\submitted{Submitted to ApJ}
% ---------------------------------------------------------------------------- %
% ---------------- ABSTRACT ---------------- %
\begin{abstract}
The stellar initial mass function (IMF) is a fundamental property in the 
measurement of stellar masses and galaxy star formation histories. In this work 
we focus on the most massive galaxies in the nearby universe 
$\log(M_{\star}/M_{\odot})>11.2$.  We obtain high quality Magellan/LDSS-3 long 
slit spectroscopy with a wide wavelength coverage of $0.4\mu{\rm m}-1.01\mu{\rm m}$ for 
41 early-type galaxies (ETGs) in the MASSIVE survey, and derive high S/N spectra 
within an aperture of $R_{\rm e}/8$. Using detailed stellar synthesis models, we 
constrain the elemental abundances and stellar IMF of each galaxy through full 
spectral modeling.  All the ETGs in our sample have an IMF that is steeper than 
a Milky Way (Kroupa) IMF.  The best-fit IMF mismatch parameter, 
$\alpha_{\rm IMF}=(M/L)/(M/L)_{\rm MW}$, ranges from 1.12 to 3.05, with an average 
of $\langle \alpha_{\rm IMF} \rangle=1.84$, suggesting that on average, the IMF is 
more bottom-heavy than Salpeter.  Comparing the estimated stellar mass with the 
dynamical mass, we find that most galaxies have stellar masses smaller than 
their dynamical masses within the $1\sigma$ uncertainty.  We complement our 
sample with lower-mass galaxies from the literature, and confirm that 
$\log(\alpha_{\rm IMF})$ is positively correlated with $\log(\sigma)$, 
$\log(M_{\star})$, and $\log(M_{\rm dyn})$. The IMF in the centers of more 
massive ETGs is more bottom-heavy. In addition, we find that $\log(\alpha_{\rm IMF})$ 
is positively correlated with both [Mg/Fe] and the estimated total metallicity 
[Z/H]. We find suggestive evidence that the effective stellar surface density 
$\Sigma_{\rm Kroupa}$ might be responsible for the variation of $\alpha_{\rm IMF}$. 
We conclude that $\sigma$, [Mg/Fe] and [Z/H] are the primary drivers of the 
global stellar IMF variation.  
\end{abstract} 

\maketitle

% ---------------------------------------------------------------------------- %
\section{Introduction}

The stellar initial mass function (IMF) describes the distribution of stellar 
masses at birth in one star formation event, and is a crucial element in astrophysical 
studies on multiple scales, from the formation of planetary systems to star formation 
and stellar feedback, from galaxy evolution to the dark matter content.  However, 
constraining the IMF is not easy since it is not directly measurable \citep{Kroupa2013}. 
Over the past few decades, studies of Galactic field stars and stellar clusters reveal 
that the stellar IMF within the Milky Way (MW) has little variation and is usually 
described as the canonical IMF \citep{Scalo1986, Kroupa2001, Bastian2010}.  The 
universal canonical IMF is often adopted as a basic assumption in galaxy modeling 
and interpretation of observational galaxy properties.

The idea of a universal IMF has been challenged by many studies.  For early 
type galaxies (ETGs), this topic have been explored by several independent 
methods \citep{Smith2020}, including stellar dynamical modeling \citep[e.g.][]{Schwarzschild1979, 
Thomas2011, Dutton2012, Cappellari2013, Li2017, Liepold2020, McConnell2012}, 
strong gravitational lensing \citep[e.g.][]{Spiniello2011, Treu2010, 
Newman2017}, and stellar population synthesis (SPS) \citep[e.g.][]{Cenarro2003, 
vanDokkum2010, vanDokkum2012, Conroy2012b, Villaume2017, Conroy2017}. 
In particular, the SPS method relies on the strength of absorption features in the 
optical to near-infrared (NIR) that are sensitive to surface gravity \citep{Wing1969}, 
as they contain information on the relative fraction of giant and dwarf stars. 
Current SPS models make use of empirical stellar libraries and theoretical response 
functions to disentangle various chemical abundances from the stellar IMF 
\citep{Cenarro2003, Vazdekis2016} and constrain the IMF in the low-mass regime 
($\leq1M_{\odot}$).  Usually two approaches are adopted in SPS: 
the full spectral modeling technique \citep[e.g.][]{Conroy2012b, Conroy2017} or 
spectral indices analysis \citep[e.g.][]{LaBarbera2019, Martin-Navarro2021}. 

SPS studies \citep[e.g.][]{Spiniello2012, Ferreras2013, Conroy2012b}, 
gravitational lensing studies \citep[e.g.][]{Treu2010} and dynamical modeling 
\citep[e.g.][]{Lasker2013, Cappellari2013} all reveal the trend that the IMF 
becomes increasingly bottom heavy with increasing velocity dispersion in ETGs.  
The agreement across different methods provides confirmation 
that the general trend of IMF variation is correct.  On the other hand, on 
an object-by-object basis, the consistency is not necessarily as strong as 
expected, as shown in \citet[e.g.][]{Smith2014, Newman2017}.

Many studies in recent years have explored possible physical mechanisms behind IMF 
variation. In addition to the correlation between the IMF and galaxy stellar velocity 
dispersion, galaxy properties such as mass density \citep[e.g.][]{Spiniello2015}, 
and stellar populations such as stellar metallicity 
\citep[e.g.][]{Martin-Navarro2015, vanDokkum2017, Parikh2018},  
[Mg/Fe] \citep[e.g.][]{Conroy2012, Smith2012}, and age \citep[e.g.][]{Barbosa2021} 
have been suggested to be correlated with IMF variation.  There is still ongoing 
debate about which galaxy properties or physical mechanism is the 
primary driver of IMF variation.

In this work, we perform a detailed full spectral modeling analysis of the most 
massive galaxies in the nearby universe to determine their stellar population parameters 
and stellar IMF. We apply a state-of-the-art stellar synthesis modeling tool to high 
quality Magellan/LDSS3 optical-NIR spectra.  
\footnote{This paper includes data gathered with the 6.5 meter Magellan Telescopes 
located at Las Campanas Observatory, Chile.}  
The sample is selected from the volume-limited MASSIVE survey (D$<108$~Mpc,
$\log(M_{\star}/_{\odot}>11.5$). We focus on a narrow mass and velocity dispersion range, 
and look for correlations between galaxy stellar populations, dynamical properties and 
the stellar IMF.  The goal of this paper is to investigate their central properties 
(within one eighth of the effective radius $R_{\rm e}$).  In particular, we combine our sample 
with lower-mass galaxies from \citet{Conroy2012} and \citet{Cappellari2013}, and present 
the global scaling relations between the IMF and velocity dispersion ($\sigma$), 
stellar mass, and dynamical mass.  
We examine the relation between the IMF and other galaxy properties in order   
to find the physical properties responsible for the scatter 
at fixed velocity dispersion.  
Furthermore, we test the consistency between our results and dynamical 
constraints.

In \S~2 we summarize the MASSIVE sample, the comparison samples, the 
observations, and the data reduction procedures. In \S~3 we summarize 
the spectral modeling tool.  In \S~4 we present our main results, including 
the central stellar population properties, global scaling relations, 
and the drivers of IMF variations 
among galaxies.  In \S~5 we assess the consistency between the derived stellar 
IMF and dynamical constraints.  We discuss results from different models and 
the physical implications of our results in \S~6 .

% ---------------------------------------------------------------------------- %
\section{Data}
\subsection{Sample}
% ---- observation ---- %
Targets in this work are selected from the MASSIVE survey \citep{Ma2014}.  
This is a volume-limited sample of the 116 most massive galaxies within 108~Mpc.  
The survey is already equipped with spatially resolved stellar kinematics from the 
Mitchell/VIRUS-P Integral Field Spectrograph (IFS) at the McDonald Observatory
in $3650$--$5850$\AA~\citep{Veale2017a, Veale2017b, Veale2018}, $V$ and $K$-band 
photometry from 2MASS (absolute $K$-band magnitude $M_K<-25.3$~mag) and wide-field 
and deep $K$-band photometry from CFHT.  Galaxies in the sample reside in a wide 
range of environments \citep{Veale2017b} from isolated to massive galaxy clusters.
Therefore the MASSIVE survey contains the ideal sample for a comprehensive study 
of very high mass galaxies.  In this work, we observe 41 ETGs in the MASSIVE survey 
using the LDSS3, selected based on their declination. These observations significantly 
extend the S/N and wavelength coverage of the existing spectroscopy.

\subsection{Comparison Samples}

The goal of the paper is to study the stellar initial mass function within and 
among massive ETGs.  We focus on a relatively narrow range of stellar mass and 
$\sigma$ and aim to look for IMF variations with galaxy properties.  Another main 
goal is to study the scatter in the IMF in galaxies with similar $\sigma$.  To 
study any global trends involving the IMF, we must also compare our results with 
lower mass ETGs.  In many previous studies of the MASSIVE sample, the comparable 
ATLAS$^{\rm 3D}$ project \citep[][hereafter A3D]{Cappellari2011}, a volume limited 
survey within 42~Mpc, has been used as a comparison sample at the low mass end 
\citep{Ma2014, Davis2016, Veale2017b}.  In this work, in order to study relations 
between IMF and galaxy properties, we seek a comparison sample where similar stellar 
population synthesis modeling exists over a comparable physical aperture.

Our main comparison sample comes from \citet[][hereafter CvD]{Conroy2012} which consists 
of 34 galaxies in the SAURON sample \citep{Bacon2001} and 4 galaxies in the Virgo cluster. 
As will be discussed in \S~3, there are several differences in detailed methodology 
between this paper and CvD. However, both works focus on the central region of 
ETGs within an effective circular radius of $R_{\rm e}/8$, and the $M/L$ and $\alpha_{\rm IMF}$ 
are constrained by the same spectral modeling tool.  Therefore, the 38 ETGs in CvD are 
a natural comparison sample at lower stellar mass.  The IMF measurements for 
ATLAS$^{\rm 3D}$ galaxies \citep[e.g.][]{Cappellari2013} also constrain the $M/L$ 
and IMF in lower-mass ETGs, but through dynamical modeling, making CvD a more natural 
comparison sample. We use CvD to explore scaling relations between the IMF and 
galaxy properties.  We also examine dynamical masses in \S~5 and discuss the importance 
of matched apertures in comparing with dynamical masses that are necessarily measured 
over much larger spatial scale, given the apparently steep gradients in low-mass IMF slope.  
We adopt the A3D sample and dynamical masses from \citet{Cappellari2013} for the 
latter discussion. 

% ---------------------------------------------------------------------------- %
\begin{figure*}[t]
\vskip 0.15cm
\centerline{\psfig{file=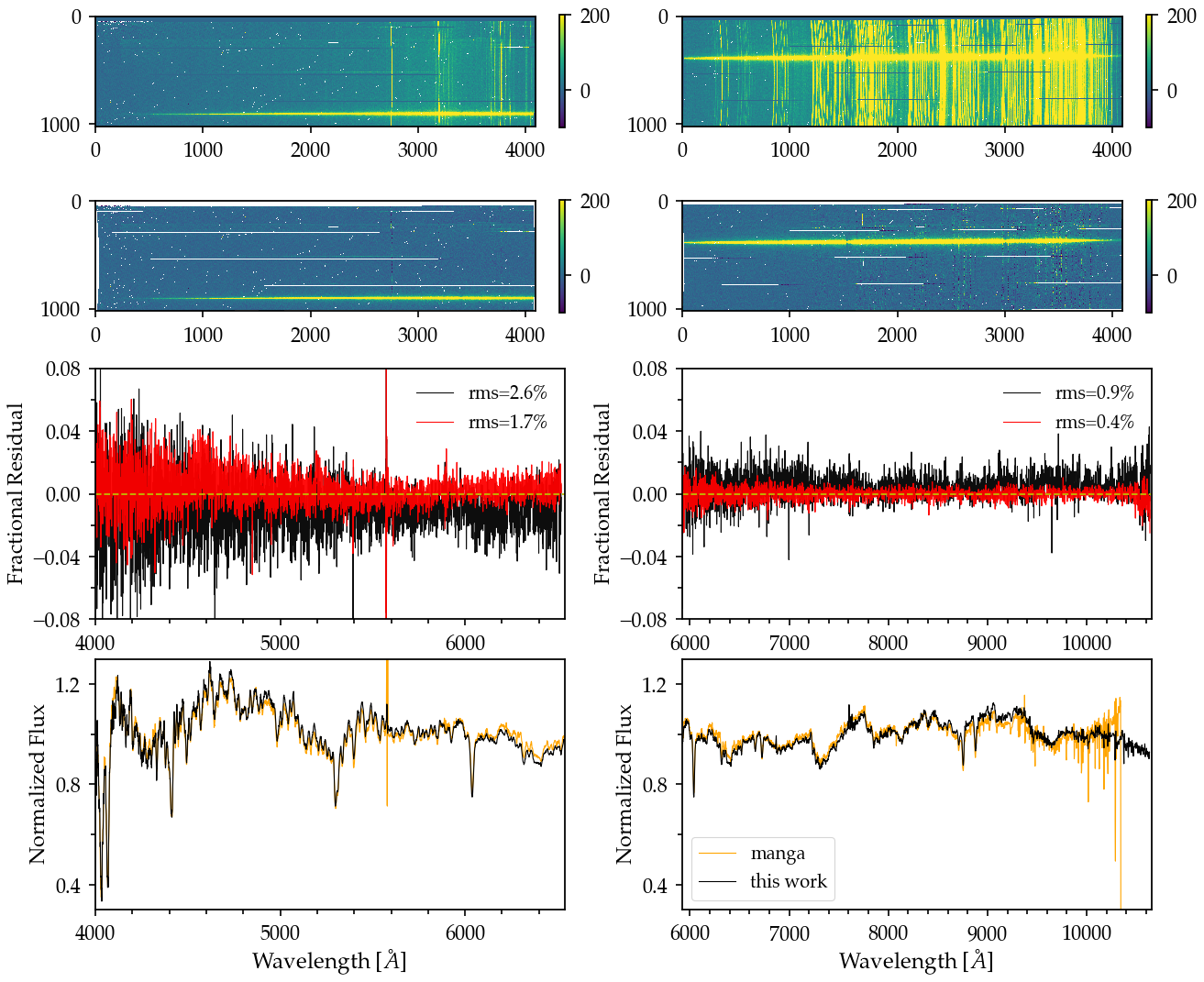, width=18cm}}
\caption{
An illustration of the sky-subtraction performance with NGC~4839 as an example.  
Left and right panels show results from the blue and red spectrographs, respectively.  
Top two panels show the 2D images before and after sky subtraction.   
In the red, most sky emission lines are removed.  The fractional sky-minus-sky 
residuals of a single exposure (black) and all exposures in one night (red, 
including four exposures in the blue and six exposures in red) are 
shown in the the third row.  The rms of fractional residual are marked at 
the top right corner.
In the bottom panels, the flux-normalized spectrum of NGC~4839 assuming a 
circular aperture of $R_{\rm e}/4$ is compared to a stacked spectrum in the same aperture 
observed by MaNGA.  The fluxes are consistent  with the MaNGA results and the 
residuals in the regions contaminated by sky emission lines are mostly clean. 
 }
\label{fig_skysub}
\end{figure*}
\noindent 
% ---------------------------------------------------------------------------- %

\subsection{Data Collection}

We present deep spectroscopy from Magellan/LDSS3 observations for 41 MASSIVE galaxies. 
The wide wavelength coverage ($4000-10300$\AA) and high signal-to-noise (S/N) of our spectra 
enable measurements of kinematics, elemental abundances, age, and stellar IMF.  
The average S/N at $4500-5500$\AA~and $8000-9000$\AA~are 
$\langle S/N\rangle=124$\AA~$^{-1}$ and $=234$\AA$^{-1}$ respectively.

Our targets were observed in eleven nights from August 2016 to March 2017 with 
the 1.0\asec wide long slit on the LDSS-3 instrument at the Magellan Clay telescope: 
March 7th-9th, 2016, August 24th-26th, 2016 and March 22nd - 26th, 2017.  
The spectral resolution is $R=1425$ and $1358$ in VPH-Blue and Red.  We derive 
the wavelength-dependent instrumental resolution from measurements of arc and sky 
lines to find, on average, $\sigma=81$ and $82$\kms in VPH-Blue and Red.
The wavelength coverage is 3800--6200\AA~and 6000--10000\AA~in the VPH-Blue and 
VPH-Red grisms, respectively.  A blocking filter, OG590, is included during the 
observation with the VPH-red grism to eliminate second order contamination at 
$\lambda>7000$\AA.  Observations are taken in the $2\times 1$ binning mode, resulting 
in a pixel scale of 0.378~\asec/pixel. The typical exposure times are $3\times 800$s 
in VPH-Blue, and $4\times 600$s with the VPH-Red grism for each target.
The slit was not always oriented along the galaxy photometric major axis.  We 
adopt the position angle measured in the Siena Galaxy Atlas Survey 
\citep{Moustakas2021} and calculate the corresponding effective radius at the observed 
slit angle to ensure that the apertures cover the same 
fraction of $R_{\rm e}$ for all galaxies.  Flat and arc exposures are taken in the afternoon 
prior to each observation night, and standard stars and bias exposures are taken during 
each observation night. All observations are read out in ``fast'' mode, with a readout 
noise of 3.5$e^{-1}$. 

We adopt distances for 19 galaxies in our sample based on infrared surface brightness 
fluctuation measurement from \citet{Jensen2021}. In addition, we utilize galaxy properties 
measured in previous MASSIVE Survey papers \citep{Ma2014, Veale2018, Ene2020}, 
including distances for other galaxies in the sample, foreground galactic extinction, and 
$K$-band photometry. For most targets, we adopt effective radii ($R_{\rm e}$) 
in semi-major axis and $K$-band luminosity measured from observations with WIRCam on the 
Canada France Hawaii Telescope (CFHT) \citep[Quenneville et al., in prep][]{}.  
In brief, stacked images were produced using the {\tt WIRWolf} pipeline \citep{Gwyn2014} 
and elliptical isophotes were determined using {\tt ARCHANGEL}.  The contaminating 
sources are defined as pixels more than four standard deviations away from the mean of 
the intensity and are excluded.  The intensity and its uncertainty are measured in each 
isophote, and the sky level is determined from the curve of growth. The half-light radii 
are measured as the radii containing half of the total light through interpolation of the 
curve of growth.  

The average ratio between $R_{\rm e}$ measured in \citep[Quenneville et al., in prep][]{} 
and adopted in previous works in the MASSIVE survey \citep[][]{Ma2014} (from 2MASS) is 1.17.
For the remaining seven galaxies, we adopt the $R_{\rm e}$ used in previous MASSIVE papers, 
but multiple the radius by this ratio to keep consistency through the sample.  On average 
the $R_{\rm e}$ along the semi-major axis of our sample is $\langle R_{\rm e} \rangle=7.2\pm2.8$~kpc, 
where the quoted error refers to the root-mean-square deviation.  The calculation of SDSS-$r$ 
band luminosity is based on $R_{\rm e}$ along the semi-major axis. We utilize the ellipse-fitting surface 
brightness profiles from the Siena Galaxy Atlas Survey \citep{Moustakas2021} in the DESI 
Legacy Imaging Surveys where available, 
or SDSS {\tt cmodel} $r-$band magnitude \citep{Gunn1998,Abolfathi2017}.

\subsection{Data Reduction}
% ---- overview ---- %
We use an updated version of the data reduction pipeline by \citet{Newman2017}.  This 
version of the pipeline has been translated into Python and has major updates in wavelength 
calibration and sky subtraction.  We briefly summarize the steps as follows and highlight 
the differences.  

% ---- ccd combine, bias, dark ---- %
Raw data from the observations are stored in two files in FITS format from two 
independent amplifiers.  The bias level is measured through the overscan regions,  
and is subtracted from the two frames separately.  Data are then converted from ADU 
to electrons units using a gain of $1.67e^-/{\rm ADU}$ and $1.43e^-/{\rm ADU}$ in 
the two amplifiers respectively.  After combining two frames, we get FITS 
files each containing an image of 4096$\times 1024$ pixels in the wavelength 
and spatial directions.  The median of the bias exposures is subtracted 
from each calibration and science exposure.

% ---- arc, wavelength calibration---- %
Wavelength solutions are derived from the He, Ne, and Ar lamp exposures in both the blue and 
red cameras observed prior to each night of observation.  In addition to the nightly 
lamp arc lines, we also make use of the sky lines from one science exposure by excluding 
$\pm70$ pixels from the galaxy center in the spatial direction, so that the wavelength range 
redder than $9000$\AA~ is better sampled by the sky lines.  For all exposures used for 
wavelength calibration, a background estimate based on the 2nd percentile of all pixel 
values in the image is subtracted. From the spectrum in the center, a set of bright lines 
are identified and matched with the arc (Ne, He, Ar) or sky-line locations.  The lines 
are usually not aligned in the spatial direction. Therefore, the identification was first 
performed on the spectrum located in the center.  For all other locations in the spatial 
direction, we take the curvature into account and allow a small overall offset of all line 
centers (no larger than 5 pixels) included in the fitting. Every 10 columns in the red and 
5 columns in the blue are median combined and used to sample in the spatial direction.  
In this way typically we are able to locate 22 arc lines 
in each arc exposure, and 40 sky lines from science exposures in the red.  
For the 2D fitting, we follow steps similar to {\ttfamily{PypeIt}} 
\citep{pypeit:joss_pub,pypeit:zenodo}.  
After locating the arc and sky lines at different spatial 
locations, we perform 2D Legendre-polynomial fitting.  For the arc lines 
in the blue camera we use orders of 6 and 3 in the wavelength and spatial 
direction, respectively.  For the red camera, we combine the arc and sky emission 
line measurements and fit them with orders of 6 in both the wavelength 
and spatial direction.  This step gives us a two-dimensional wavelength solution map.  

In addition, spectra are usually not perfectly aligned parallel to the CCD pixels.  
To trace the spectral distortion, we add two dithered exposures of the same galaxy 
to map the distortion perpendicular to the dispersion direction.  In a pair of dithered 
exposures, galaxies are usually not located in the center, but in the top and bottom half 
of the CCD.  In the combined image, we trace the locations of the two galaxy peaks 
every 5 pixels throughout the spatial directions.  We then build a trace map by fitting 
3rd and 5th order polynomials in the blue and the red, respectively. The above steps provide us 
two sets of coefficient arrays for the wavelength and trace solutions.

% ---- flat normalization ---- %
We took dome flat exposures using Quartz lamps in the afternoon prior to the science 
observations.  A two-dimensional master flat for each night is constructed using the 
median of all flat exposures.  We then model the lamp spectrum and divide that out to 
isolate the pixel-to-pixel sensitivity variations and the non-uniformity in the slit.  
To model the lamp spectrum, we first fit for a slit function which shows the variation along 
the slit by performing an iterative one-dimensional 4th order b-spline fit to the median 
spatial profile of the central half pixels in the wavelength direction.  With the spatial 
profile divided out, we fit for a one-dimensional lamp spectrum using the median spectrum 
in the central half pixels in the spatial direction using a cubic b-spline fit.  We then construct a 
two-dimensional lamp spectrum using the wavelength solution. Dark pixels are identified 
and marked in this step.  The final flat frame for normalization is derived by dividing 
the median flat by the two-dimensional model of the lamp spectra and then normalized by 
the median value. 
% ---- cosmic ray ---- %
All science exposures are normalized by the final flat frame. Cosmic rays 
are identified and removed by {\tt L.A.Cosmic} \citep{vanDokkum2012}. 
For each science exposure, the small wavelength zero-point correction is 
derived by fitting for the differences in a few sky line at all columns 
($5577.338$\AA~, $6300.304$\AA~in the blue, and $6300.304$\AA~, $9375.961$\AA~, 
$10418.363$\AA~in the red)

% ---------------------------------------------------------------------------- %
\begin{figure*}[ht]
\vskip 0.15cm
  \centerline{\psfig{file=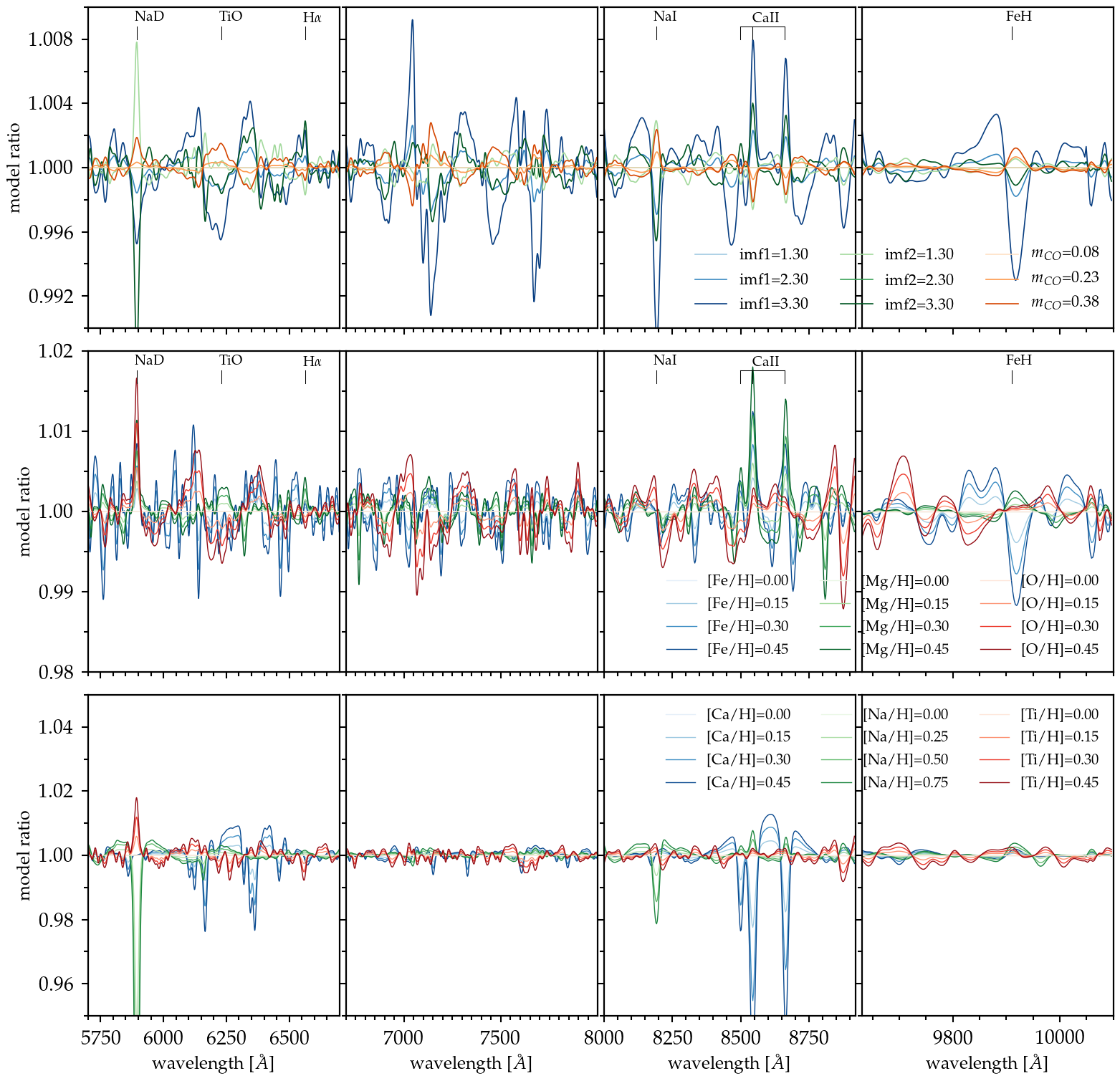, width=18cm}}
  \caption{Illustration of model sensitivity to the change of 
  stellar population parameters as a function of wavelength.  Models are smoothed to 
  $\sigma=250~km\;s^{-1}$.  In all panels the reference model has solar abundances 
  and a Kroupa IMF.  The ratio between each model and the reference are normalized 
  by a polynomial with one order per 100\AA.  In the top panels, {\tt imf1} and 
  {\tt imf2} represent the IMF slopes below and above $0.5M_{\odot}$ and $m_{co}$ 
  represents the low mass cutoff.  The top panels show the flux-normalized ratio 
  between each model with parameters varied relative to 
  the model with a Kroupa IMF ({\tt imf1}=1.3, {\tt imf2}=2.3, {\tt imf3}=0.08). 
  The middle and bottom panels show the regions that are sensitive to 
  changes in the abundances of Fe, Mg, O, Ca, Na and Ti.  
Although in some regions it looks like an increase in {\tt imf1} 
  and {\tt imf3} result in similar change of model ratio, e.g., NaD, TiO, NaI and CaT, 
  the changes among different features (e.g. NaI, CaT and the Wind Ford band) 
  have slight differences and in principle it can break their degeneracy.
}
\label{fig_alfmodel}
\end{figure*}
% ---------------------------------------------------------------------------- %
\begin{figure*}[t]
  \centerline{\psfig{file=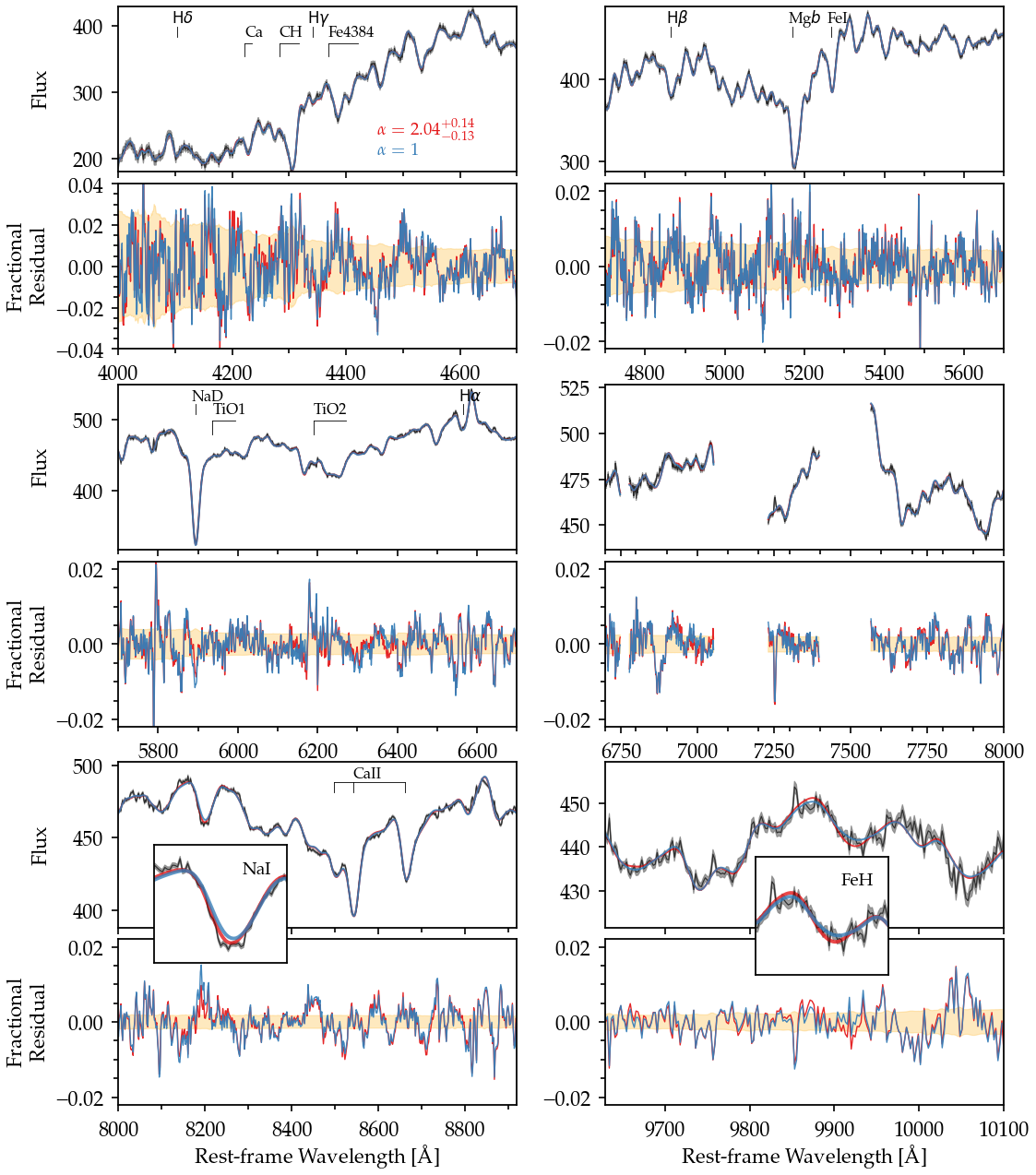, width=17cm}}
  \caption{Comparison between two models of the spectrum of NGC~5490 
  extracted from an effective circular aperture of $R_{\rm e}/8$. Data (black) 
  and model flux are shown in the top panels, and the fractional residuals in 
  the bottom panels.  Our fiducial model has a flexible IMF with three parameters 
  (low and high mass slopes, and a low mass cutoff) and is shown in red. 
  A model with a fixed Kroupa IMF is shown in blue.  The two models result in 
  a ratio of IMF mismatch parameter of 2.04. Our best model reveals an IMF 
  more bottom heavy than Salpeter.  Note that the residuals from the fiducial 
  model are noticeably smaller in the IMF sensitive features, especially around 
  NaI, NaD, TiO, Mg$b$ and the Wing-Ford band. 
 }
\label{fig_n5490}
\end{figure*}
% ---------------------------------------------------------------------------- %

% ---- sky subtraction ---- %
We follow sky subtraction procedures similar to the MUSE \citep{Weilbacher2020}
and MaNGA \citep{Bundy2015} data reduction pipelines \citep{Law2016}.  The regions 
used to fit for sky spectra are chosen to be $\pm250$ pixels away from galaxy centers, except 
for three very nearby galaxies in the Virgo cluster where we increase the threshold 
to $\pm350$ pixels.  From the Siena photometry, the average $r$-band surface brightness 
at 250 and 350 pixels for all galaxies are $24.8$ and $26.4$~\sb.
We divide the sky regions into continuum and sky emission line regions, and those 
in the sky emission line regions are further divided into several groups based on 
the origin of the sky emission lines following \citet{Weilbacher2020}.  The reason 
is that the sky emission lines scale similarly within the same group. Skylines with 
flux above $0.5\%$ of the brightest sky line are considered in the scaling.  
Among these sky lines, we start with the group with the brightest total flux, and assume 
pixels within $3$\AA~ from the emission line center belong to the group.  The remaining pixels 
are assumed to be in the continuum region. A relative scaling factor array is 
then derived by comparing the $4\sigma$-clipped mean flux of all pixels in the same group in 
different exposures. The scale factor array is normalized by the minimum 
before applied to all exposures. 

To model the sky background for one exposure, we use the sky regions in the exposure 
itself and its two neighboring exposures. The sky is then modeled using a procedure 
similar to \citep{Law2016}: due to the wavelength correction in each exposure, the 
sky pixels in 3 exposures sample the sky spectra much better than a single exposure.  
We smooth the inverse variance spectra by a box car of width 100~pixels in the 
continuum, and 3 pixels within 3\AA~of emission lines.  This step helps prevent a bias 
towards lower values in the sky modeling.  Then a sky model is constructed with an 
iterative two-dimensional b-spline function, with break points spacing of 1.0 pixel 
in the blue, and 0.7 in the red, and a third order polynomial in the spatial 
direction.  The sky model for each exposure is then multiplied by the scaling factor 
in each emission line group and the continuum region.  
		
In Figure ~\ref{fig_skysub}, we show our sky subtraction performance in one exposure of
NGC~4839 in the red.  Left and right panels show results in the blue and red 
spectrographs, respectively. The top two panels show the 2D exposures before and after 
sky subtraction, while the middle panel shows the 2D sky model.
Note that in the red, most of the sky emission is removed.  
The third panel shows the fractional sky residuals as a function of wavelength.  
We randomly select 200 residual spectra in the sky regions (regions that are $\geq$250 
pixels away from galaxy centers), and plot the median residual 
as a fraction of the median sky spectra.  The fractional residuals are less 
than $5\%$ at all wavelengths.  In the bottom panels, the flux-normalized spectrum 
of NGC~4839 assuming a circular aperture of $R_{\rm e}/4$ is compared to a 
stacked spectrum in the same aperture observed by MaNGA DR8.  
The flux is consistent with the MaNGA results and the residuals in the regions 
contaminated by sky emission lines are low. 
% ---- rectification ---- %
Sky models and sky subtracted exposures and the corresponding inverse variance array 
are all rectified based on the wavelength and spatial grids. Spectra are first linearly 
resampled with flux conserved to a common linear wavelength grid of 4096 pixels 
starting at 3774\AA~ and 5923\AA~ in the blue and red, respectively, and width of 
0.6757\AA~ and 1.156\AA.  They are then resampled onto a common grid in the spatial direction 
based on the trace solution.  We note that there is a significant amount of scattered light 
in the far red with LDSS-3.  This is primarily indicated by the increasingly extended 
spatial profile around 1~$\mu$m.  The scattered light cannot be perfectly 
removed and pixels near the Wing-Ford band are often contaminated.

\subsection{Extraction}

We use effective circular radii of $R_{\rm e}/8$ (\S~2.2). To mimic the circular aperture, 
we assign different weights to pixels as a function of distance from the galaxy center, 
$r$. Specifically, all pixels within 0.5\asec~ have weights of unity and outer pixels 
are assigned with weight of $\pi \times r$.  Standard stars are extracted in a radius of 4.0\asec. 

The standard stars used in this work are EG274, GD71 and LTT7987. To perform flux calibration, each standard star spectrum from the VizieR archive is divided from our standard star observations. 
Then the galaxy spectrum is normalized by the smoothed calibration spectrum.
During spectral extraction, we assign weights to pixels in the spatial 
direction to mimic a circular aperture. 
The telluric correction is calculated for extracted galaxy spectra.
We model and remove telluric absorption features by fitting atmospheric models 
using the {\tt MOLECFIT} code, assuming the contributing molecular species are 
H$_2$O and O$_2$.  The bands used for telluric modeling are $6250-6300$\AA in the 
blue, and $6820-6970$\AA, $7210-7330$\AA, $7590-7690$\AA, $8170-8360$\AA~and 
$9100-9400$\AA\ in the red.  The telluric correction derived is then used for other apertures.

% ---------------------------------------------------------------------------- %
% -- Table1 -- %
\begin{table*}[t]
\caption{Key Properties and Fitting Results of MASSIVE galaxies}
\renewcommand{\arraystretch}{1.85}
%\centering
\resizebox{0.875\width}{!}{\begin{tabular}{lcccccccccccccc}
\toprule
Name & $R_{\rm e}$ & $\log{(L_r)}$ & $\log{(M_{\rm dyn})}$ & $\log{(M_{\star})}$ 
& S/N &$\sigma$ & [Fe/H] & [Mg/Fe] & $\log{\rm (age)}$ & [O/Fe] &  [Na/Fe] &  $M/L_r$ &  $M/L_i$ &  $\alpha_{\rm IMF}$ \\
 & [$^{\prime\prime}$] & [$L_{\odot}$] & [$M_{\odot}$] & [$M_{\odot}$] & [\AA$^{-1}$] & [\kms]& & & [Gyr] & &  &  [$M_{\odot}/L_{\odot}$] &  [$M_{\odot}/L_{\odot}$] &  \\
(1) & (2) & (3) & (4) & (5) & (6) & (7) & (8) & (9) & (10) & (11) & (12) & (13)& (14) & (15)\\
\hline
NGC~0057 & 17.07 &10.86 & $11.58^{+0.09}_{-0.11}$ & $11.68^{+0.05}_{-0.05}$ & $238$ & $303^{+2}_{-2}$ & $0.12^{+0.02}_{-0.02}$ & $0.35^{+0.02}_{-0.02}$ & $1.09^{+0.03}_{-0.04}$ & $0.39^{+0.04}_{-0.04}$ & $0.47^{+0.02}_{-0.02}$ & $7.04^{+0.72}_{-0.59}$ & $4.87^{+0.48}_{-0.40}$ & $1.90^{+0.20}_{-0.15}$ \\
NGC~0080 & 21.05 &11.00 & $11.66^{+0.07}_{-0.07}$ & $11.82^{+0.06}_{-0.07}$ & $226$ & $240^{+2}_{-1}$ & $0.11^{+0.01}_{-0.01}$ & $0.29^{+0.02}_{-0.02}$ & $0.91^{+0.08}_{-0.04}$ & $0.32^{+0.04}_{-0.04}$ & $0.41^{+0.02}_{-0.03}$ & $5.76^{+0.94}_{-0.62}$ & $4.07^{+0.66}_{-0.43}$ & $1.93^{+0.39}_{-0.23}$ \\
NGC~0533 & 31.56 &11.05 & $11.75^{+0.07}_{-0.07}$ & $11.90^{+0.05}_{-0.06}$ & $195$ & $250^{+2}_{-2}$ & $0.05^{+0.01}_{-0.02}$ & $0.33^{+0.01}_{-0.02}$ & $1.11^{+0.03}_{-0.03}$ & $0.29^{+0.03}_{-0.04}$ & $0.44^{+0.02}_{-0.02}$ & $6.46^{+0.51}_{-0.45}$ & $4.50^{+0.35}_{-0.31}$ & $1.74^{+0.15}_{-0.14}$ \\
NGC~0741 & 27.18 &11.03 & $11.81^{+0.07}_{-0.07}$ & $11.77^{+0.06}_{-0.06}$ & $279$ & $267^{+1}_{-1}$ & $0.09^{+0.01}_{-0.01}$ & $0.27^{+0.01}_{-0.01}$ & $1.10^{+0.02}_{-0.03}$ & $0.33^{+0.04}_{-0.03}$ & $0.35^{+0.01}_{-0.01}$ & $6.25^{+0.55}_{-0.45}$ & $4.38^{+0.38}_{-0.30}$ & $1.72^{+0.15}_{-0.12}$ \\
NGC~1016 & 20.53 &11.13 & $11.83^{+0.06}_{-0.07}$ & $11.83^{+0.04}_{-0.05}$ & $203$ & $288^{+2}_{-2}$ & $0.15^{+0.02}_{-0.02}$ & $0.28^{+0.02}_{-0.02}$ & $1.11^{+0.02}_{-0.03}$ & $0.16^{+0.05}_{-0.05}$ & $0.46^{+0.02}_{-0.02}$ & $6.11^{+0.61}_{-0.56}$ & $4.27^{+0.42}_{-0.38}$ & $1.54^{+0.16}_{-0.14}$ \\
NGC~1453 & 21.93 &10.89 & $11.60^{+0.10}_{-0.12}$ & $11.70^{+0.04}_{-0.05}$ & $243$ & $293^{+7}_{-2}$ & $0.05^{+0.02}_{-0.03}$ & $0.29^{+0.02}_{-0.02}$ & $1.10^{+0.04}_{-0.04}$ & $0.35^{+0.07}_{-0.05}$ & $0.45^{+0.03}_{-0.02}$ & $7.99^{+0.75}_{-0.76}$ & $5.47^{+0.48}_{-0.52}$ & $2.08^{+0.23}_{-0.24}$ \\
NGC~1600 & 29.56 &11.21 & $11.87^{+0.07}_{-0.08}$ & $12.00^{+0.06}_{-0.10}$ & $216$ & $333^{+2}_{-2}$ & $0.10^{+0.02}_{-0.02}$ & $0.31^{+0.01}_{-0.01}$ & $1.12^{+0.02}_{-0.02}$ & $0.38^{+0.04}_{-0.04}$ & $0.46^{+0.02}_{-0.02}$ & $6.55^{+0.64}_{-0.58}$ & $4.52^{+0.42}_{-0.39}$ & $1.67^{+0.16}_{-0.15}$ \\
NGC~1700 & 15.35 &10.91 & $11.31^{+0.12}_{-0.16}$ & $11.19^{+0.07}_{-0.08}$ & $341$ & $227^{+1}_{-1}$ & $0.19^{+0.02}_{-0.02}$ & $0.26^{+0.02}_{-0.02}$ & $0.64^{+0.03}_{-0.02}$ & $0.23^{+0.04}_{-0.04}$ & $0.43^{+0.02}_{-0.02}$ & $2.30^{+0.22}_{-0.20}$ & $1.72^{+0.16}_{-0.14}$ & $1.12^{+0.11}_{-0.10}$ \\
NGC~2418 & 14.02 &10.89 & $11.40^{+0.10}_{-0.12}$ & $11.62^{+0.06}_{-0.08}$ & $188$ & $234^{+2}_{-2}$ & $0.12^{+0.02}_{-0.06}$ & $0.30^{+0.02}_{-0.03}$ & $0.94^{+0.12}_{-0.04}$ & $0.28^{+0.12}_{-0.07}$ & $0.47^{+0.03}_{-0.03}$ & $5.95^{+1.18}_{-1.26}$ & $4.19^{+0.81}_{-0.90}$ & $1.89^{+0.43}_{-0.59}$ \\
NGC~2513 & 15.33 &10.92 & $11.60^{+0.08}_{-0.10}$ & $11.52^{+0.04}_{-0.05}$ & $334$ & $282^{+2}_{-2}$ & $0.08^{+0.02}_{-0.02}$ & $0.28^{+0.01}_{-0.01}$ & $1.12^{+0.02}_{-0.02}$ & $0.26^{+0.04}_{-0.03}$ & $0.43^{+0.02}_{-0.02}$ & $5.48^{+0.47}_{-0.43}$ & $3.83^{+0.33}_{-0.29}$ & $1.36^{+0.12}_{-0.12}$ \\
NGC~2672 & 20.23 &11.00 & $11.68^{+0.07}_{-0.08}$ & $11.64^{+0.06}_{-0.06}$ & $205$ & $259^{+2}_{-2}$ & $0.10^{+0.02}_{-0.01}$ & $0.30^{+0.02}_{-0.02}$ & $1.13^{+0.01}_{-0.03}$ & $0.14^{+0.04}_{-0.05}$ & $0.48^{+0.02}_{-0.02}$ & $6.80^{+0.86}_{-0.76}$ & $4.72^{+0.59}_{-0.52}$ & $1.70^{+0.22}_{-0.20}$ \\
NGC~3209 & 14.42 &10.98 & $11.63^{+0.08}_{-0.09}$ & $11.81^{+0.06}_{-0.08}$ & $206$ & $306^{+2}_{-3}$ & $0.11^{+0.02}_{-0.02}$ & $0.28^{+0.02}_{-0.02}$ & $1.07^{+0.03}_{-0.05}$ & $0.35^{+0.05}_{-0.05}$ & $0.54^{+0.03}_{-0.02}$ & $5.92^{+0.60}_{-0.63}$ & $4.12^{+0.41}_{-0.42}$ & $1.59^{+0.20}_{-0.17}$ \\
NGC~3462 & 12.37 &10.99 & $11.42^{+0.09}_{-0.11}$ & $11.53^{+0.08}_{-0.09}$ & $212$ & $221^{+1}_{-2}$ & $0.24^{+0.01}_{-0.02}$ & $0.16^{+0.02}_{-0.02}$ & $0.68^{+0.02}_{-0.03}$ & $0.03^{+0.05}_{-0.05}$ & $0.34^{+0.02}_{-0.02}$ & $4.06^{+0.63}_{-0.52}$ & $3.04^{+0.46}_{-0.38}$ & $1.98^{+0.32}_{-0.28}$ \\
NGC~3615 & 12.2 &10.98 & $11.57^{+0.09}_{-0.10}$ & $11.65^{+0.08}_{-0.08}$ & $218$ & $263^{+1}_{-1}$ & $0.10^{+0.02}_{-0.02}$ & $0.23^{+0.02}_{-0.02}$ & $1.03^{+0.05}_{-0.05}$ & $0.18^{+0.04}_{-0.03}$ & $0.33^{+0.02}_{-0.02}$ & $6.54^{+0.58}_{-0.52}$ & $4.62^{+0.40}_{-0.37}$ & $1.92^{+0.25}_{-0.20}$ \\
NGC~3805 & 11.39$^*$ &10.98 & $11.50^{+0.09}_{-0.11}$ & $11.48^{+0.04}_{-0.06}$ & $176$ & $255^{+34}_{-2}$ & $0.17^{+0.03}_{-0.02}$ & $0.31^{+0.02}_{-0.05}$ & $0.90^{+0.16}_{-0.04}$ & $0.19^{+0.06}_{-0.07}$ & $0.53^{+0.03}_{-0.10}$ & $7.55^{+1.65}_{-1.61}$ & $5.34^{+1.19}_{-1.18}$ & $2.61^{+0.67}_{-1.14}$ \\
NGC~3842 & 18.41$^*$ &10.96 & $11.67^{+0.07}_{-0.08}$ & $11.74^{+0.05}_{-0.07}$ & $183$ & $282^{+2}_{-2}$ & $0.06^{+0.02}_{-0.02}$ & $0.30^{+0.01}_{-0.01}$ & $1.12^{+0.01}_{-0.02}$ & $0.25^{+0.04}_{-0.04}$ & $0.52^{+0.02}_{-0.02}$ & $8.02^{+0.86}_{-0.64}$ & $5.53^{+0.59}_{-0.43}$ & $2.02^{+0.22}_{-0.16}$ \\
NGC~3862 & 13.62$^*$ &10.89 & $11.52^{+0.09}_{-0.10}$ & $11.69^{+0.06}_{-0.09}$ & $165$ & $259^{+15}_{-2}$ & $0.05^{+0.02}_{-0.02}$ & $0.30^{+0.02}_{-0.02}$ & $1.11^{+0.03}_{-0.05}$ & $0.21^{+0.05}_{-0.05}$ & $0.41^{+0.02}_{-0.04}$ & $9.44^{+1.42}_{-1.06}$ & $6.55^{+1.02}_{-0.72}$ & $2.54^{+0.42}_{-0.36}$ \\
NGC~3937 & 15.64$^*$ &10.98 & $11.65^{+0.07}_{-0.09}$ & $11.85^{+0.05}_{-0.04}$ & $210$ & $282^{+2}_{-2}$ & $0.06^{+0.02}_{-0.03}$ & $0.26^{+0.02}_{-0.02}$ & $1.11^{+0.02}_{-0.04}$ & $0.13^{+0.07}_{-0.05}$ & $0.37^{+0.02}_{-0.02}$ & $6.96^{+1.10}_{-0.65}$ & $4.92^{+0.76}_{-0.44}$ & $2.00^{+0.28}_{-0.19}$ \\
NGC~4055 & 10.18$^*$ &10.89$^*$ & $11.53^{+0.09}_{-0.11}$ & $11.59^{+0.03}_{-0.05}$ & $134$ & $281^{+4}_{-10}$ & $0.09^{+0.03}_{-0.03}$ & $0.34^{+0.03}_{-0.03}$ & $1.09^{+0.04}_{-0.06}$ & $-0.00^{+0.11}_{-0.11}$ & $0.22^{+0.05}_{-0.04}$ & $6.00^{+1.29}_{-1.32}$ & $4.32^{+0.88}_{-0.94}$ & $1.59^{+0.40}_{-0.41}$ \\
NGC~4073 & 31.07$^*$ &11.16 & $11.95^{+0.05}_{-0.06}$ & $11.88^{+0.06}_{-0.06}$ & $224$ & $296^{+2}_{-2}$ & $0.08^{+0.02}_{-0.02}$ & $0.31^{+0.02}_{-0.02}$ & $1.11^{+0.03}_{-0.03}$ & $0.20^{+0.04}_{-0.04}$ & $0.44^{+0.02}_{-0.02}$ & $6.68^{+0.60}_{-0.55}$ & $4.66^{+0.41}_{-0.37}$ & $1.75^{+0.15}_{-0.15}$ \\
NGC~4472 & 66.43$^*$ &10.99 & $11.62^{+0.09}_{-0.11}$ & $11.60^{+0.03}_{-0.03}$ & $396$ & $290^{+1}_{-1}$ & $0.15^{+0.01}_{-0.01}$ & $0.27^{+0.01}_{-0.01}$ & $0.98^{+0.01}_{-0.01}$ & $0.25^{+0.03}_{-0.02}$ & $0.44^{+0.01}_{-0.01}$ & $5.07^{+0.23}_{-0.21}$ & $3.55^{+0.15}_{-0.14}$ & $1.49^{+0.08}_{-0.07}$ \\
NGC~4486 & 75.74$^*$ &10.82 & $11.45^{+0.12}_{-0.15}$ & $11.55^{+0.04}_{-0.05}$ & $276$ & $327^{+4}_{-4}$ & $-0.00^{+0.03}_{-0.03}$ & $0.39^{+0.02}_{-0.02}$ & $1.13^{+0.01}_{-0.05}$ & $0.38^{+0.05}_{-0.05}$ & $0.57^{+0.03}_{-0.02}$ & $6.70^{+0.75}_{-0.55}$ & $4.56^{+0.51}_{-0.35}$ & $1.58^{+0.18}_{-0.15}$ \\
NGC~4555 & 12.8 &11.08 & $11.72^{+0.08}_{-0.09}$ & $11.74^{+0.07}_{-0.11}$ & $216$ & $317^{+2}_{-2}$ & $0.12^{+0.02}_{-0.02}$ & $0.31^{+0.02}_{-0.02}$ & $0.86^{+0.04}_{-0.05}$ & $0.17^{+0.05}_{-0.05}$ & $0.45^{+0.02}_{-0.02}$ & $8.26^{+0.73}_{-0.88}$ & $5.82^{+0.51}_{-0.62}$ & $3.05^{+0.39}_{-0.36}$ \\
NGC~4649 & 62.59$^*$ &10.82 & $11.48^{+0.12}_{-0.16}$ & $11.43^{+0.03}_{-0.04}$ & $400$ & $330^{+2}_{-2}$ & $0.12^{+0.01}_{-0.01}$ & $0.31^{+0.01}_{-0.01}$ & $1.14^{+0.00}_{-0.01}$ & $0.30^{+0.04}_{-0.04}$ & $0.71^{+0.02}_{-0.02}$ & $6.69^{+0.40}_{-0.42}$ & $4.58^{+0.27}_{-0.28}$ & $1.54^{+0.09}_{-0.10}$ \\
NGC~4839 & 30.64$^*$ &10.97 & $11.85^{+0.06}_{-0.07}$ & $11.65^{+0.08}_{-0.12}$ & $230$ & $257^{+2}_{-2}$ & $0.03^{+0.02}_{-0.02}$ & $0.35^{+0.02}_{-0.02}$ & $1.09^{+0.03}_{-0.04}$ & $0.08^{+0.05}_{-0.04}$ & $0.36^{+0.02}_{-0.02}$ & $5.26^{+0.56}_{-0.56}$ & $3.73^{+0.39}_{-0.39}$ & $1.49^{+0.18}_{-0.15}$ \\
NGC~4874 & 38.82 &11.23 & $12.02^{+0.05}_{-0.06}$ & $11.69^{+0.07}_{-0.09}$ & $127$ & $244^{+2}_{-2}$ & $0.04^{+0.02}_{-0.02}$ & $0.29^{+0.02}_{-0.02}$ & $1.09^{+0.03}_{-0.05}$ & $0.23^{+0.05}_{-0.05}$ & $0.32^{+0.02}_{-0.02}$ & $4.96^{+0.59}_{-0.53}$ & $3.51^{+0.41}_{-0.36}$ & $1.40^{+0.18}_{-0.15}$ \\
NGC~5129 & 17.33 &11.11 & $11.74^{+0.06}_{-0.07}$ & $11.65^{+0.08}_{-0.11}$ & $182$ & $249^{+2}_{-1}$ & $0.06^{+0.02}_{-0.01}$ & $0.27^{+0.02}_{-0.02}$ & $0.87^{+0.05}_{-0.05}$ & $0.26^{+0.05}_{-0.05}$ & $0.30^{+0.02}_{-0.02}$ & $3.47^{+0.52}_{-0.44}$ & $2.52^{+0.37}_{-0.30}$ & $1.29^{+0.20}_{-0.17}$ \\
NGC~5208 & 10.7 &10.85 & $11.73^{+0.12}_{-0.16}$ & $11.59^{+0.10}_{-0.19}$ & $190$ & $253^{+3}_{-3}$ & $0.20^{+0.03}_{-0.05}$ & $0.25^{+0.03}_{-0.03}$ & $0.76^{+0.15}_{-0.07}$ & $0.05^{+0.10}_{-0.08}$ & $0.58^{+0.04}_{-0.03}$ & $6.67^{+0.93}_{-1.23}$ & $4.81^{+0.67}_{-0.87}$ & $2.72^{+0.52}_{-0.74}$ \\
NGC~5490 & 10.59 &10.81 & $11.49^{+0.12}_{-0.16}$ & $11.49^{+0.05}_{-0.08}$ & $296$ & $344^{+2}_{-2}$ & $0.11^{+0.01}_{-0.02}$ & $0.31^{+0.01}_{-0.01}$ & $1.10^{+0.03}_{-0.03}$ & $0.34^{+0.04}_{-0.04}$ & $0.56^{+0.02}_{-0.02}$ & $8.12^{+0.45}_{-0.48}$ & $5.52^{+0.29}_{-0.32}$ & $2.04^{+0.14}_{-0.13}$ \\
NGC~6375 & 12.28 &10.90 & $11.42^{+0.09}_{-0.10}$ & $11.66^{+0.07}_{-0.07}$ & $205$ & $223^{+4}_{-2}$ & $0.16^{+0.02}_{-0.02}$ & $0.24^{+0.02}_{-0.02}$ & $1.01^{+0.10}_{-0.06}$ & $0.21^{+0.04}_{-0.05}$ & $0.37^{+0.03}_{-0.04}$ & $8.45^{+0.94}_{-0.86}$ & $5.82^{+0.67}_{-0.60}$ & $2.47^{+0.52}_{-0.40}$ \\
NGC~6442 & 12.69 &10.92 & $11.44^{+0.08}_{-0.10}$ & $11.67^{+0.04}_{-0.05}$ & $229$ & $239^{+2}_{-2}$ & $0.07^{+0.02}_{-0.01}$ & $0.23^{+0.01}_{-0.01}$ & $1.11^{+0.02}_{-0.03}$ & $0.20^{+0.04}_{-0.04}$ & $0.32^{+0.02}_{-0.02}$ & $6.19^{+0.55}_{-0.52}$ & $4.32^{+0.37}_{-0.35}$ & $1.66^{+0.15}_{-0.14}$ \\
NGC~6482 & 13.74 &10.78 & $11.43^{+0.13}_{-0.18}$ & $11.50^{+0.02}_{-0.05}$ & $295$ & $269^{+3}_{-2}$ & $0.11^{+0.02}_{-0.02}$ & $0.38^{+0.03}_{-0.02}$ & $0.98^{+0.04}_{-0.03}$ & $0.42^{+0.03}_{-0.04}$ & $0.64^{+0.04}_{-0.04}$ & $9.27^{+0.91}_{-1.28}$ & $6.34^{+0.62}_{-0.86}$ & $2.83^{+0.42}_{-0.44}$ \\
NGC~7052 & 21.61 & $-$ & $11.72^{+0.07}_{-0.09}$ & $-$ & $214$ & $276^{+2}_{-2}$ & $0.06^{+0.02}_{-0.02}$ & $0.31^{+0.02}_{-0.02}$ & $1.11^{+0.03}_{-0.03}$ & $0.39^{+0.05}_{-0.05}$ & $0.40^{+0.02}_{-0.02}$ & $6.67^{+0.73}_{-0.62}$ & $4.63^{+0.49}_{-0.43}$ & $1.82^{+0.19}_{-0.19}$ \\
NGC~7619 & 22.75 &10.89 & $11.61^{+0.09}_{-0.11}$ & $11.49^{+0.04}_{-0.06}$ & $322$ & $313^{+2}_{-2}$ & $0.14^{+0.01}_{-0.01}$ & $0.31^{+0.02}_{-0.02}$ & $0.99^{+0.03}_{-0.03}$ & $0.29^{+0.03}_{-0.04}$ & $0.53^{+0.02}_{-0.02}$ & $5.74^{+0.41}_{-0.38}$ & $3.99^{+0.28}_{-0.26}$ & $1.65^{+0.15}_{-0.12}$ \\
NGC~7626 & 23.05 &10.94 & $11.56^{+0.09}_{-0.10}$ & $11.50^{+0.04}_{-0.06}$ & $303$ & $261^{+1}_{-1}$ & $0.07^{+0.01}_{-0.01}$ & $0.35^{+0.01}_{-0.01}$ & $1.06^{+0.02}_{-0.03}$ & $0.35^{+0.03}_{-0.03}$ & $0.53^{+0.02}_{-0.02}$ & $4.97^{+0.32}_{-0.31}$ & $3.46^{+0.22}_{-0.21}$ & $1.31^{+0.10}_{-0.08}$ \\
UGC~10918 & 19.27 & $-$ & $11.74^{+0.07}_{-0.07}$ & $-$ & $209$ & $267^{+3}_{-3}$ & $0.06^{+0.03}_{-0.04}$ & $0.29^{+0.02}_{-0.03}$ & $1.06^{+0.05}_{-0.11}$ & $0.39^{+0.06}_{-0.06}$ & $0.38^{+0.03}_{-0.03}$ & $6.73^{+0.77}_{-0.79}$ & $4.73^{+0.52}_{-0.53}$ & $2.06^{+0.35}_{-0.27}$ \\
NGC~7436 & 20.58 &11.02 & $11.88^{+0.06}_{-0.07}$ & $11.82^{+0.07}_{-0.09}$ & $237$ & $289^{+2}_{-2}$ & $0.13^{+0.02}_{-0.02}$ & $0.29^{+0.02}_{-0.02}$ & $1.12^{+0.02}_{-0.03}$ & $0.26^{+0.04}_{-0.04}$ & $0.53^{+0.03}_{-0.02}$ & $7.44^{+0.57}_{-0.56}$ & $5.14^{+0.38}_{-0.36}$ & $1.89^{+0.16}_{-0.13}$ \\
NGC~7556 & 24.99 &11.05 & $11.78^{+0.07}_{-0.07}$ & $11.71^{+0.08}_{-0.13}$ & $166$ & $232^{+2}_{-2}$ & $0.06^{+0.01}_{-0.02}$ & $0.29^{+0.01}_{-0.01}$ & $1.11^{+0.02}_{-0.03}$ & $0.15^{+0.04}_{-0.04}$ & $0.40^{+0.02}_{-0.02}$ & $5.60^{+0.58}_{-0.58}$ & $3.92^{+0.40}_{-0.40}$ & $1.48^{+0.17}_{-0.15}$ \\
NGC~7386 & 16.25 &10.94 & $11.73^{+0.07}_{-0.09}$ & $11.51^{+0.07}_{-0.12}$ & $225$ & $291^{+2}_{-2}$ & $0.13^{+0.02}_{-0.02}$ & $0.30^{+0.02}_{-0.02}$ & $1.01^{+0.05}_{-0.07}$ & $0.14^{+0.04}_{-0.05}$ & $0.48^{+0.03}_{-0.03}$ & $6.11^{+0.58}_{-0.61}$ & $4.32^{+0.40}_{-0.43}$ & $1.77^{+0.23}_{-0.20}$ \\
NGC~0547 & 30.51$^*$ &10.73$^*$ & $11.83^{+0.06}_{-0.07}$ & $11.44^{+0.03}_{-0.06}$ & $242$ & $234^{+1}_{-1}$ & $0.07^{+0.02}_{-0.02}$ & $0.31^{+0.02}_{-0.01}$ & $1.08^{+0.03}_{-0.08}$ & $0.35^{+0.04}_{-0.04}$ & $0.36^{+0.03}_{-0.02}$ & $6.02^{+0.43}_{-0.43}$ & $4.20^{+0.31}_{-0.29}$ & $1.68^{+0.19}_{-0.12}$ \\
NGC~0545 & 27.06$^*$ &10.91$^*$ & $12.11^{+0.06}_{-0.06}$ & $11.49^{+0.08}_{-0.14}$ & $233$ & $224^{+2}_{-1}$ & $0.08^{+0.01}_{-0.02}$ & $0.28^{+0.02}_{-0.01}$ & $1.10^{+0.03}_{-0.04}$ & $0.26^{+0.04}_{-0.03}$ & $0.30^{+0.02}_{-0.02}$ & $4.74^{+0.31}_{-0.30}$ & $3.37^{+0.21}_{-0.21}$ & $1.42^{+0.09}_{-0.09}$ \\
\hline
\end{tabular}}
\tablecomments{
Columns are (1): Galaxy name. 
(2) Effective radii ($R_{\rm e}$) along semi-major axis measured from CFHT-K band images 
where available, or $R_{\rm e}$ from 2MASS multiplied by the adopted ratio of 1.17 
(marked by $^*$, see \S~2.3).
(3) $r-$band luminosity based on the Siena Atlas photometry where 
available, or from SDSS (marked by $^*$). 
(4) Dynamical mass estimated based on empirical relation from \citet{Cappellari2006}.
(5) Stellar mass estimated based on extrapolated luminosity weighted $M/L_r$ within 
$R_{\rm e}$ and $r-$band luminosity.
(6) mean S/N per \AA~in $8000-9000$\AA.
(7-15) {\tt alf} fitting results from spectra stacked within $R_{\rm e}/8$ 
in an effective circular aperture.}
%\vspace{0.1cm}
\end{table*}
% ---------------------------------------------------------------------------- %
\begin{figure*}[t]
\centerline{\psfig{file=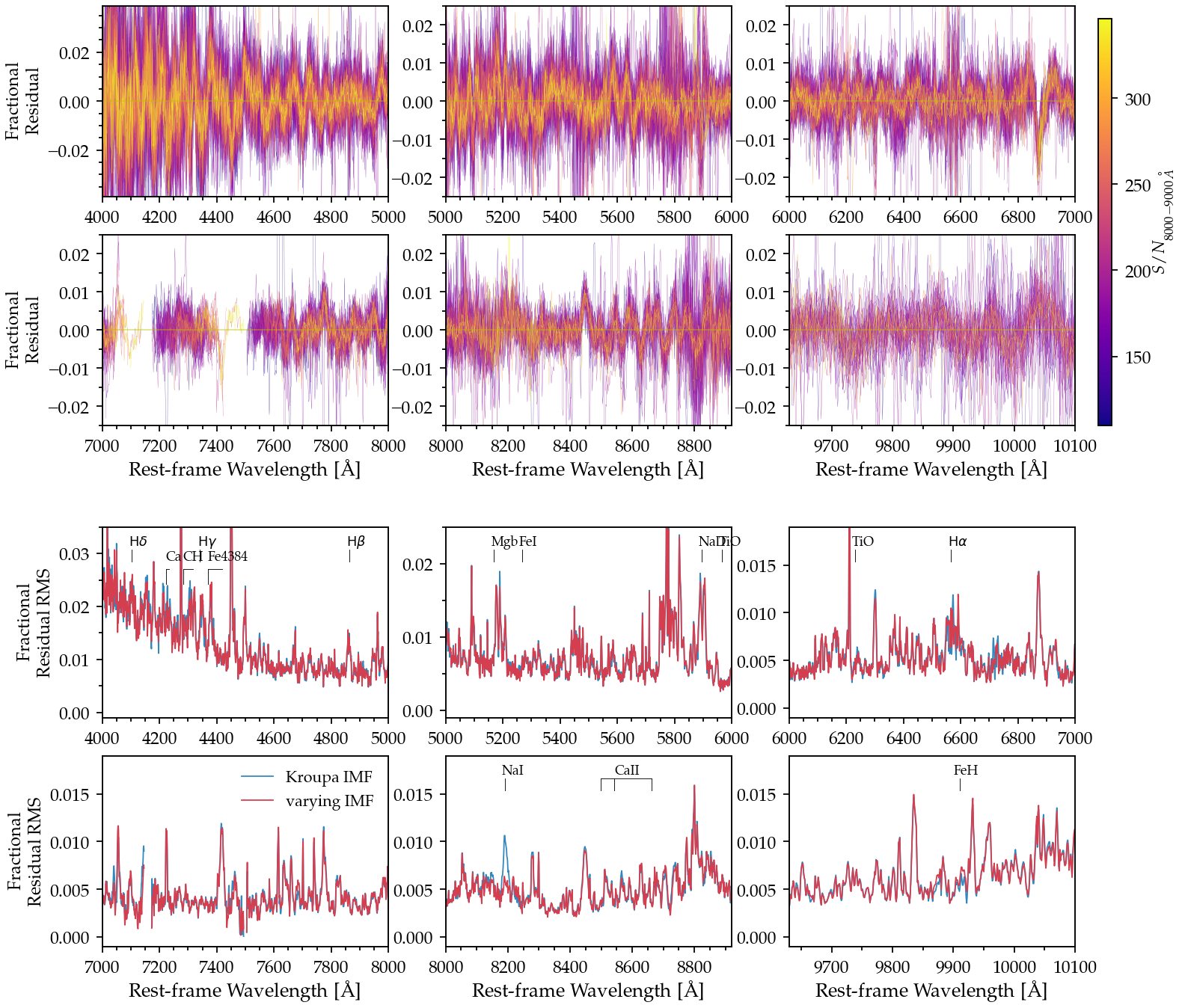, width=18cm}}
\caption[Two numerical solutions]{
Top: Fractional residuals from the fiducial IMF model 
(two power-law IMF + low cutoff mass) for all galaxies in our sample.  
Colors indicate the mean S/N in $8000-9000$\AA.  On average the fractional 
residuals are smaller than $2\%$. For spectra with S/N $>300$, the fraction 
residual is better than $1\%$ in the red.
Bottom: A comparison of fractional residual rms of all 41 galaxies from 
fitting with a fixed Kroupa IMF, and the fiducial model.  A model with a fixed 
Kroupa IMF creates much larger residuals in several regions, specifically around 
NaI.}
\label{fig_residual}
\end{figure*}
\noindent  
% ---------------------------------------------------------------------------- %

\section{Spectral Fitting}

We model the spectra using the absorption line fitter \citep[{\tt alf},][]{Conroy2012, 
Conroy2014, Conroy2018}.  We present the details of the modeling tool in \S~3.1, then 
show the best-fit models and residuals, along with a comparison to fixed-IMF results, 
in \S~3.2.

\subsection{Modeling}
{\tt alf} enables full spectral stellar population modeling for stellar ages $>1$Gyr 
and for metallicities from $\sim-2.0$ to $+0.25$.  The Markov Chain Monte Carlo (MCMC) algorithm \citep[{\tt emcee},][]{ForemanMackey2013} is used in the exploration of 
parameter space.  
Currently {\tt alf} adopts the MIST stellar isochrones \citep{Choi2016} and utilizes 
a new spectral library \citep{Villaume2017} that includes continuous wavelength coverage 
from $0.35-2.4\mu m$.  It also utilizes the theoretical response functions that were 
computed using the ATLAS and SYNTHE programs \citep{Kurucz1970, Kurucz1993}.  The 
theoretical response functions tabulate the effect on the spectrum of enhancing each of 
18 individual elements.  

There are 36 free parameters involved in the fitting: galaxy kinematics (radial 
velocity, and velocity dispersion), a two-burst star formation history (two ages 
and a mass fraction), and stellar populations, including the overall metallicity 
([Z/H]) and 18 individual element abundances (Fe, Mg, O, C, N, Na, Si, K, Ca, Ti, 
V, Cr, Mn, Co, Ni, Sr, Ba, Eu). In the two-burst model, the young component has as a 
prior an upper age limit of 3~Gyr, while the older component prior has an upper limit of 
14~Gyr. Both have a lower limit on their age of 0.5~Gyr. We adopt flat priors from 
$500-10500$~\kms~for recession velocity, $10-1000$~\kms~for velocity dispersion, and 
$-1.8$--$0.3$ for [Fe/H].  The priors are zero outside these ranges. The prior ranges 
for elemental abundances are $-0.3-0.5$ except for [Na/H], which is $-0.3-1.0$. 

There are several possible forms for the stellar IMF in {\tt alf}.  In our fiducial 
model, we use {\tt alf} to fit three parameters for the stellar IMF: two logarithmic 
slopes in $dn/dm\propto{M^{-\gamma}}$, in the mass ranges $0.08M_{\odot}<M<0.5M_{\odot}$ 
({\tt imf1}), $0.5M_{\odot}<M<1.0M_{\odot}$ ({\tt imf2}), and a low mass cutoff ({\tt imf3}). 
At $M\geq1.0M_{\odot}$, the slope is fixed to the value of a Salpeter slope, $\gamma = 2.3$. The 
priors on {\tt imf1} and {\tt imf2} are both flat with ranges of $0.5\leq$imf1/imf2$\leq3.5$, 
and $0.08M_{\odot}\leq$imf3$\leq0.4M_{\odot}$. A Kroupa IMF with {\tt imf1}=1.3 and 
{\tt imf2}=2.3 is used as a reference.  From the $M/L$ posterior, the IMF mismatch parameter, 
$\alpha_{\rm IMF}\equiv(M/L)/(M/L)_{\rm Kroupa}$, is also calculated.  This parameter refers to 
the best-ft $M/L$ normalized by the $M/L$ based on the reference IMF, and indicates how 
much the best-fit IMF deviates from the reference IMF. Since $M/L$ is sensitive 
not only to the IMF, but also age, metallicity, etc, this normalization is helpful since $\alpha_{\rm IMF}$ 
is only sensitive to the IMF. In this work the Salpeter IMF is 
assumed to be in the form of a single power law with a slope of 2.35 over a mass range 
of $0.1$--$100M_{\odot}$, which is different from what is defined in \citet{Salpeter1955}.
The IMF mismatch parameter of the Salpeter IMF is assumed to be $\alpha_{\rm IMF}\approx1.55$.

Several ``nuisance'' parameters are included: hot star components, emission lines 
(H, [O II], [O III], [S II], [N I], and [N II]), and an error jitter term to correct 
the observational uncertainties.  For each spectrum, the continuum is first normalized by 
a polynomial fitted to the ratio between model and data.  The polynomial has an order of $(\lambda_{max}$--$\lambda_{min})/100$\AA.  We fit in five separate wavelength intervals: 
$0.4-0.47$,$0.47-0.57$,$0.57-0.67$,$0.675-0.80$,$0.80-0.892$ and $0.963-1.01\mu m$.
Pixels with strong telluric contamination in the following ranges are masked out in 
the observed frames: $0.717-0.735\mu m$, $0.686-0.689\mu m$ and $0.752-0.769\mu m$.

% parameter sensitivity
Figure ~\ref{fig_alfmodel} illustrates the sensitivity of the model to the change in 
parameters, as a function of wavelength.  Models with varying parameters are compared to 
a reference model with solar stellar populations and a Kroupa IMF.  All models are smoothed 
to $\sigma=250 \, \mathrm{km s^{-1}}$ in this figure.  The ratio between each model and 
the reference are normalized by a polynomial with an order of 
$(\lambda_{\rm max}$--$\lambda_{\rm min})/100$\AA. In the top panels, {\tt imf1} and {\tt imf2} 
represent the IMF slopes below and above $0.5M_{\odot}$ and $m_{\rm co}$ represents the 
low mass cutoff.  Figure ~\ref{fig_alfmodel} illustrates the relative wavelength sensitivity 
of a few parameters.  For example, an increase in {\tt imf1}, {\tt imf2} or [Na/H] 
will be reflected in a deeper NaI feature, suggesting that the NaI feature is sensitive 
to more than just the Na abundance. In addition, the change caused by a varying low-mass 
cutoff({\tt imf3}) at different wavelengths has a different amplitude compared to the changes 
caused by a varying IMF slope at $0.08-0.5M_{\odot}$ ({\tt imf1}), or $0.5-1.0M_{\odot}$ 
({\tt imf2}), indicating that the effect due to {\tt imf3} cannot be simply compensated 
by {\tt imf1} or {\tt imf2}. 

Since the goal of this work is to study global variations in the IMF and their
connection with stellar populations, CvD is our main comparison sample (\S~2.2).  
Although we are using the same method as CvD, there are 
some differences in the details and we list them as follows:
%Here we list several difference in methodology from CvD.  
We use a version of {\tt alf} with updated stellar libraries \citep{Villaume2017}, 
stellar isochrones \citep{Choi2016} and response functions. Details of the updates 
are summarized in \citet{Conroy2017}.  Also, we include wavelength intervals of 
$0.57-0.67 \mu{\rm m}$ and $0.67-0.80 \mu{\rm m}$, which are not included in CvD.  As shown 
in Figure ~\ref{fig_alfmodel}, the additional wavelength intervals in our work indicate 
that we have more information to constrain the IMF and the abundances of elements 
such as Mg, O, Ti.  We do not include the nuisance parameters {\tt logm7g} and 
{\tt teff} as free parameters in the fitting. {\tt logm7g} and {\tt teff} represent 
the light at $1\mu{\rm m}$ contributed by an M7III giant star, and the shift in $T_{\rm eff}$ 
relative to the fiducial isochrones, respectively.  They were introduced for systematics 
testing in an early version of {\tt alf} when the isochrones were not metallicity 
dependent and are not useful in the current version. 
{\tt logtrans} represents the strength of the atmospheric transmission 
function for H$_2$O and O$_2$ together.  Since we have corrected for the atmospheric 
transmission function from H$_2$O and O$_2$ separately using $\tt MOLECFIT$, {\tt logtrans} 
is not included in the fit.  These differences make our model setup more comparable 
to \citet{Newman2017}.

\subsection{Best-fit Models and Residuals}

%We first use one galaxy, NGC~5490, as an example to demonstrate the differences between our default IMF model and the reference Kroupa IMF model.  
In our default IMF model, the 
IMF is described by three parameters: low ({\tt imf1}) and intermediate mass ({\tt imf}) 
slopes and a low cut-off mass ({\tt imf3}).  In contrast, a Kroupa IMF has a low mass 
slope of {\tt imf1}$=1.3$ and an intermediate mass slope of {\tt imf2}=2.3.  
Since {\tt alf} uses all available pixels in the spectra, instead of single spectral 
absorption features, changes in parameters are not all visibly prominent in the residuals.  
However for galaxies in our sample, the differences between  models using the flexible 
(default) IMF or a fixed Kroupa IMF are always obvious, as fixing the IMF to the Kroupa 
form fails to describe all of the spectral features, especially the gravity-sensitive 
features. 

We use one galaxy, NGC~5490, as an example and show a comparison between the two models 
in Figure~\ref{fig_n5490}.  The spectrum has a mean $S/N=296$\AA$^{-1}$ from 
$8000-9000$\AA~in the rest-frame.  The best-fit model spectrum from our fiducial model, 
and the corresponding residuals, are shown in red.  The results from a fixed Kroupa IMF are 
shown in blue.  The $\alpha_{\rm IMF}$ from the posterior of our fiducial model 
is $\alpha_{\rm IMF}\approx2.0$, indicating that the stellar IMF in the central 
region of NGC~5490 is more bottom heavy than Salpeter.  Comparing the differences in the 
residuals, the main difference is that the residuals from our fiducial model are noticeably 
smaller in the IMF sensitive features such as NaI, NaD and FeH.  By forcing the IMF to a 
Kroupa form, {\tt alf} tries to mitigate mismatches by changing abundance ratios in 
key features. For this particular galaxy the abundances of Fe, N, Ca, Ti 
are shifted to lower values by at least $1\sigma$, and [Na/H] and $\log({\rm age})$ are shifted 
to higher values by $\sim 2\sigma$. 
NaI in particular is obviously under-fit by the Kroupa model. The [Na/H] 
abundances are $0.57$ and $0.61$ in our fiducial and Kroupa model.  
In other words, an increase in [Na/Fe] abundance is not enough to describe the strong 
NaI feature if we choose to force the IMF to be Kroupa. 
The shift in abundances by forcing the IMF to be Kroupa is consistent with what is shown 
in Figure~\ref{fig_alfmodel}, e.g., a decrease in [Ca/Fe] and increase in [Na/Fe] are 
induced to compensate the decrease of IMF slopes.
Comparing the results from forcing the IMF to Kroupa with that 
from our fiducial model for the whole sample, on average, [N/Fe], [Ti/Fe], [Ca/Fe]
are underestimated by 0.03, 0.02 and 0.02~dex, respectively. [Na/Fe] and [O/Fe] are 
overestimated by 0.02 and 0.05~dex, respectively.
Although the sensitivities of IMF parameters and elemental abundances follow similar 
trends in some regions of the spectra (e.g., from Figure~\ref{fig_alfmodel}), 
Figure~\ref{fig_n5490} demonstrates that {\tt alf} is able to break the possible 
degeneracy especially with the information from gravity-sensitive features. 

In the top sub-panel of Figure~\ref{fig_residual}, we show the fractional residuals for all 
of the galaxies in our sample using the fiducial model.  Colors indicate the median S/N of 
each spectrum at $8000-9000$~\AA.  The mean $S/N$ of all spectra in our sample is 
$\langle S/N \rangle =234$\AA$^{-1}$.  The general trend is that galaxies with higher S/N have fractional 
residuals closer to zero.   Overall the residuals are smaller than $1\%$ in the red.  
We notice that there are some wavelength dependent patterns in the residuals in the rest-frame.  
They are present in both low and high $S/N$ spectra. This indicates that such patterns 
do not depend on data reduction or telluric modeling, and are more likely caused by 
the mismatch between the galaxy spectra and the model \citep{vanDokkum2017}.

In the bottom sub figure, we compare our fiducial model with the residuals from fixing 
the IMF to Kroupa.  On average in the MASSIVE sample 
$\langle \alpha_{\rm IMF} \rangle=1.84\pm0.43\textbf{}$.  These panels show the quadratic mean of 
fractional residual over all 41 galaxies as a function of wavelength.  The residuals 
from our fiducial and reference models are shown in red and blue, respectively. 
A model with fixed Kroupa IMF results in much larger residuals in several regions: 
NaD, H$\alpha$, region around TiO features and NaI.  Despite the issue with 
scattered light in the far red, the Wing-Ford band provides an important 
constraint on both the stellar IMF and the Fe abundance so it is included 
in the fitting.  On average, the Kroupa model tends to over-estimate the 
[Na/H] and under-estimate the abundances of Fe, Ca and Ti.  

% ---------------------------------------------------------------------------- %
\begin{figure*}
  \centerline{\psfig{file=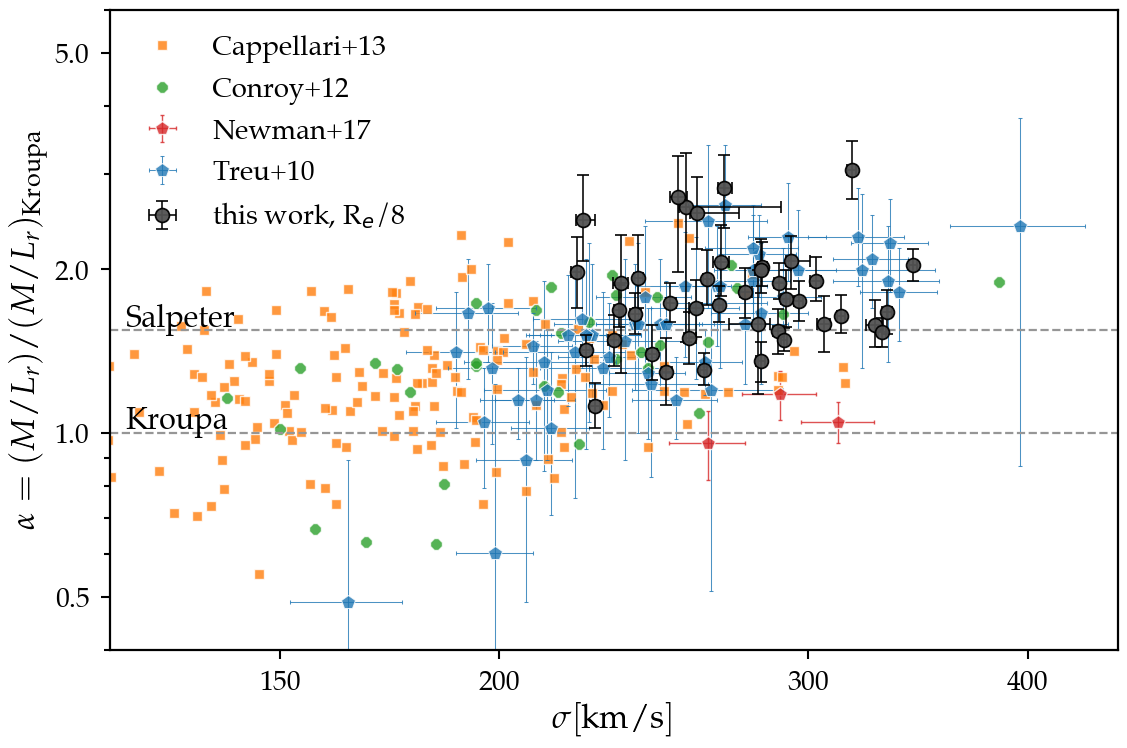, width=15cm}}
  \caption{The IMF mismatch parameter, $\alpha_{\rm IMF}$, as a function of galaxy 
  central velocity dispersion in this work (black) and previous studies based on 
  SPS (CvD), dynamical modeling (A3D), and 
  strong lensing \citep[][]{Newman2017, Treu2010}.
 }
\label{fig_alpha-sig}
\end{figure*}
% ---------------------------------------------------------------------------- %
\begin{figure*}
  \centerline{\psfig{file=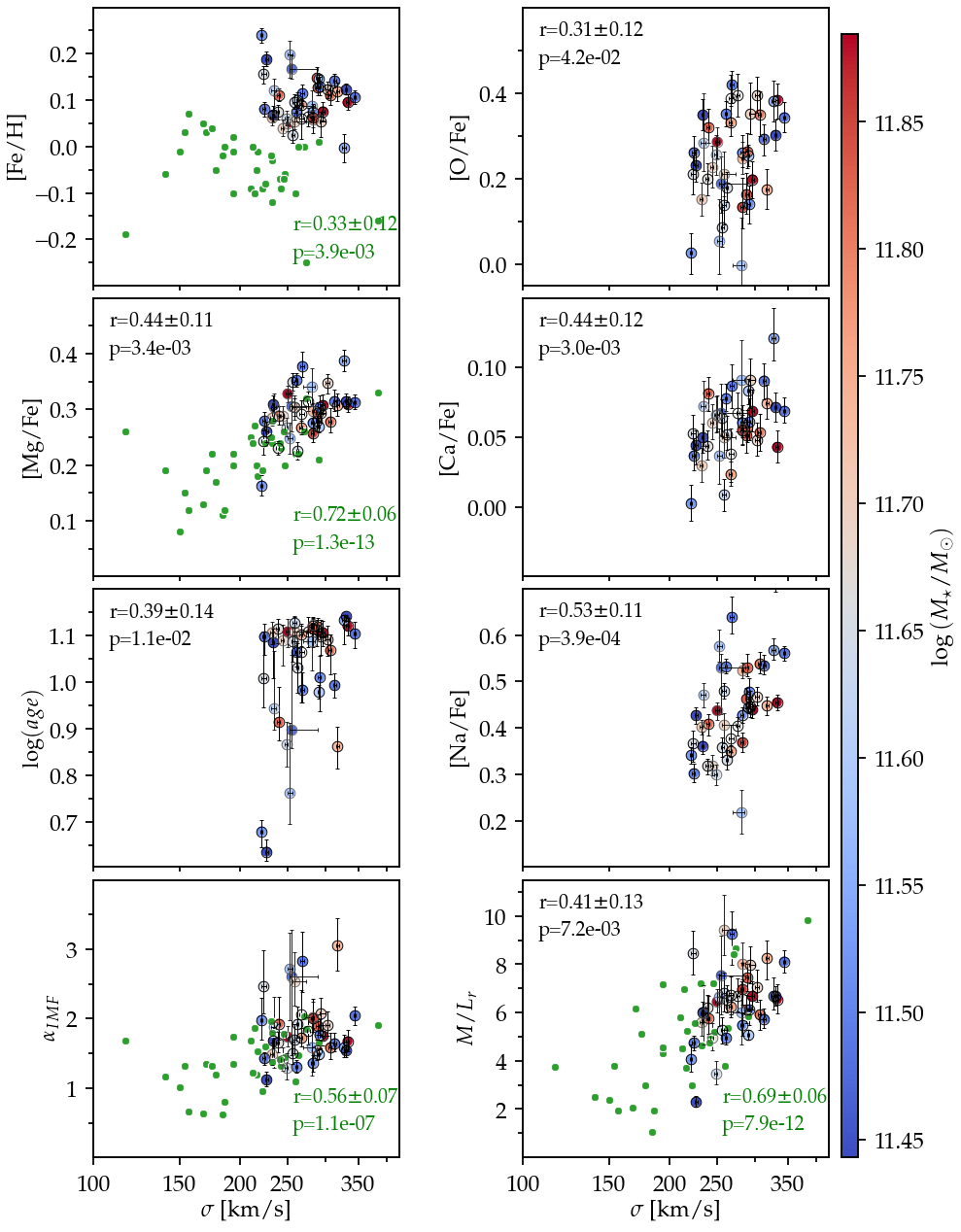, width=15cm}}
  \caption{
  From top left to bottom right: [Fe/H], [Mg/Fe], $\log({\rm age})$, IMF $\alpha-$mismatch 
  parameter, [O/Fe],  [C/Fe], [Na/Fe] and $M/L_r$ as a function of $\sigma$ for 
  all galaxies in our sample.  All properties are luminosity weighted within $R_{\rm e}$.  
  Colors indicate the stellar mass. Results from CvD (green) are included as a 
  comparison for available 
  parameters.  Pearson correlation coefficients and the 1$\sigma$ uncertainty are 
  shown when the corresponding $p-$ value is smaller than $5\%$.  
  For our sample alone and the combined sample they are shown in black and 
  green, respectively.
 }
\label{fig_sigmarel}
\end{figure*}
% ---------------------------------------------------------------------------- %
%Figure 7
\begin{figure*}[ht]
\vskip 0.15cm
  \centerline{\psfig{file=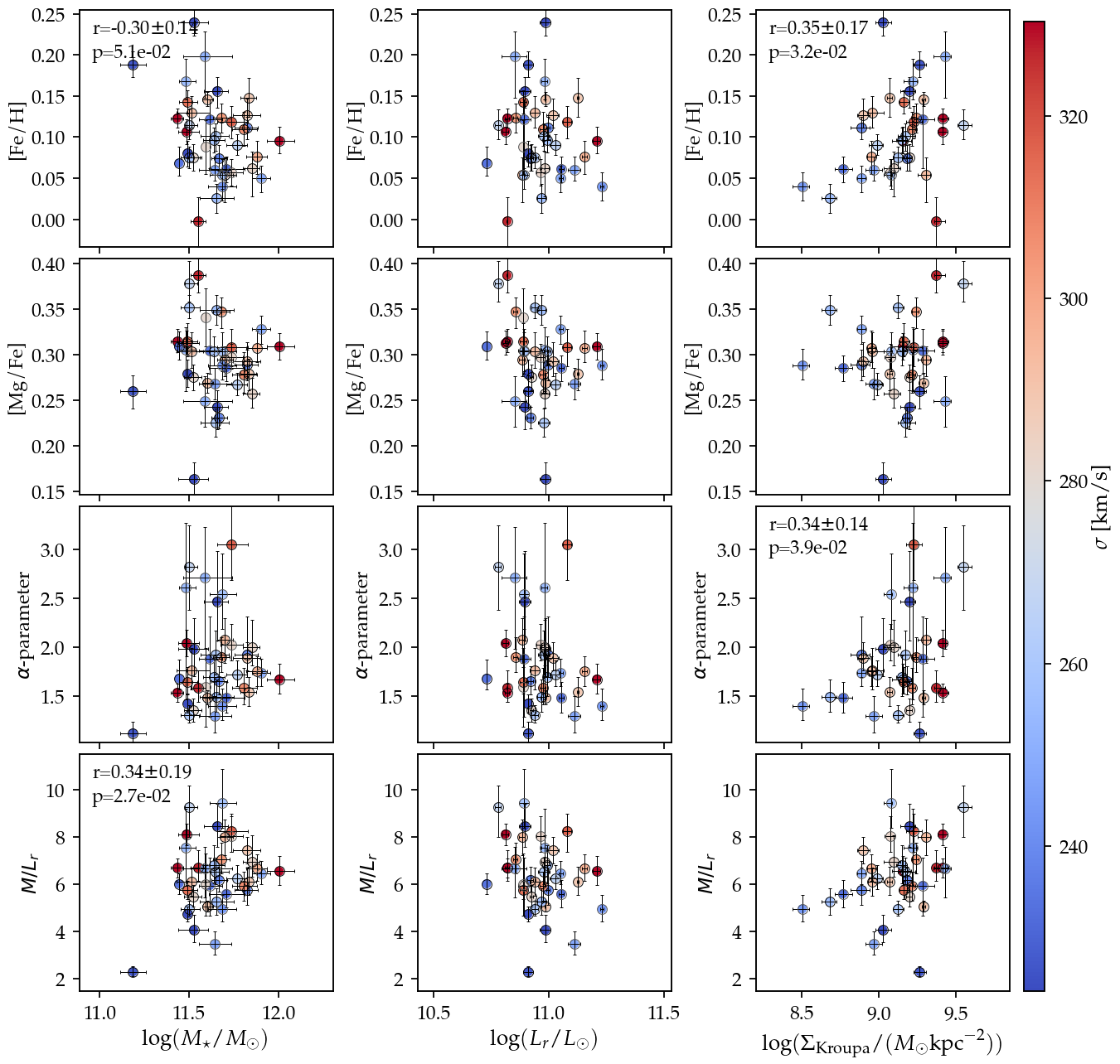, 
  width=18cm}}
  \caption{
  From top left to bottom right: [Fe/H], [Mg/Fe], IMF $\alpha-$mismatch 
  parameter and $M/L_r$ as a function of logarithmic stellar mass, 
  luminosity and effective stellar surface density for all galaxies in our sample. 
  Pearson correlation coefficients and the 1$\sigma$ uncertainty are 
  shown when the corresponding $p-$ value is smaller than $5\%$.
 }
\label{fig_mstar}
\end{figure*}
\vspace{0.1cm}
% ---------------------------------------------------------------------------- %
% ---------------------------------------------------------------------------- %
\begin{figure*}[t]
\vskip 0.15cm  \centerline{\psfig{file=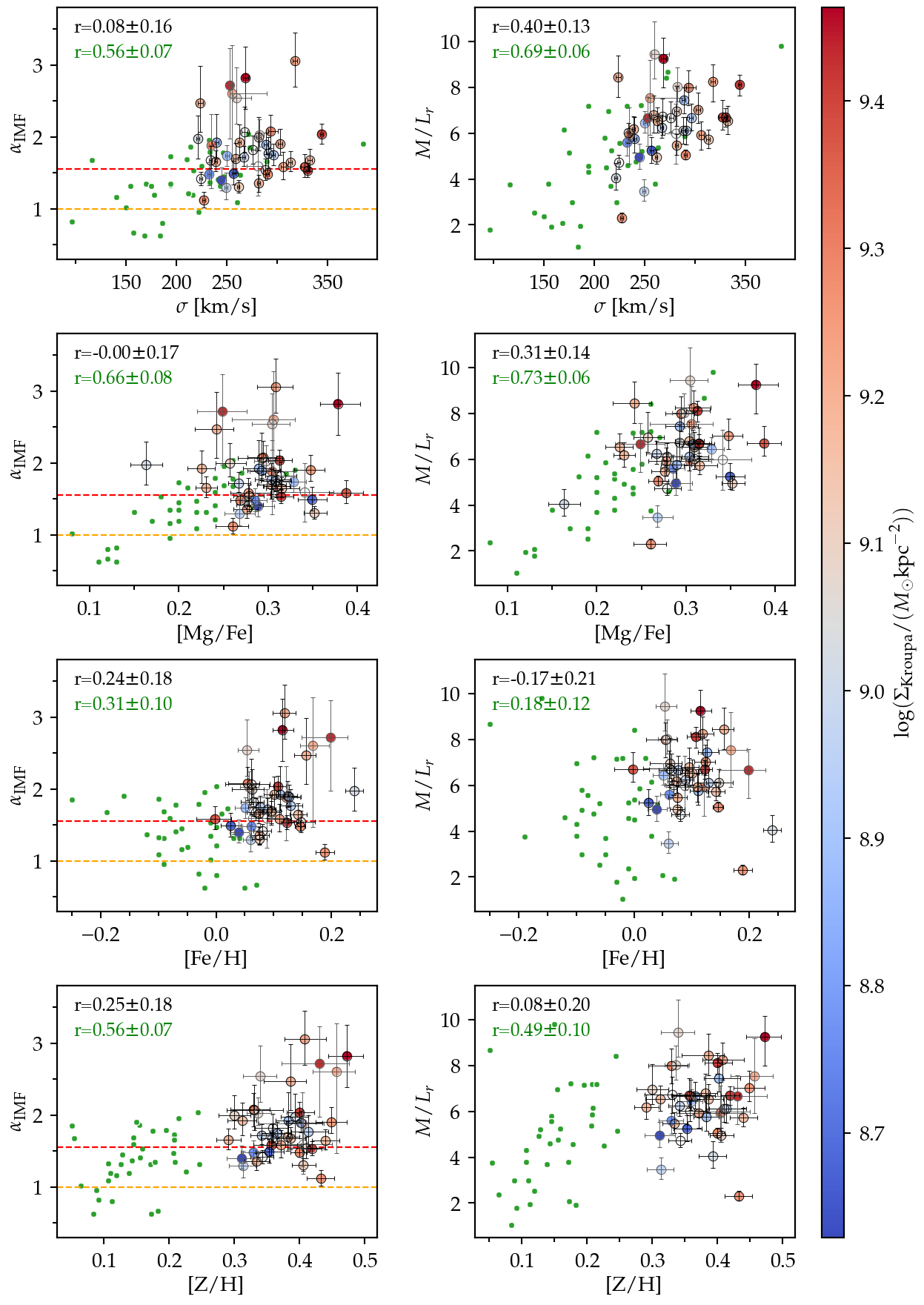, width=15cm}}
\caption{From top to bottom: $M/L_r$ and the IMF $\alpha-$mismatch 
  parameter as a function of $\sigma$, [Mg/Fe], [Fe/H] and estimated total 
  metallicity [Z/H].  Red and orange horizontal lines indicate Salpeter and 
  Kroupa IMF.  Pearson correlation coefficients and its 1$\sigma$ uncertainty 
  are shown in our sample alone (black), and in the combined sample (green).  
  The strong correlation between [Mg/Fe] and IMF suggest that the star 
  formation timescale may play an important role in shaping the stellar IMF.
}
\label{fig_alpharel}
\end{figure*}
%\noindent  
% ---------------------------------------------------------------------------- %
% ---------------------------------------------------------------------------- %

\section{Results}
We present our main results in this section.  In \S~4.1 we describe the procedures 
to derive dynamical mass, stellar masses, and the effective stellar mass density.
In \S~4.2 we present the central properties of galaxies in our sample: 
their stellar populations and stellar IMF.  We present correlations between 
stellar populations and other galaxy properties for our sample alone in \S~4.3 
and including the low-mass comparison sample in \S~4.4. 
In \S~4.5 we show multivariate linear regression involving $\alpha_{\rm IMF}$ 
and stellar populations. In \S~4.6 we briefly mention the relation with 
environment.

% ---------------------------------------------------------------------------- %
\subsection{Dynamical Mass, Stellar Mass and Stellar Surface Density}

The dynamical masses are calculated based on an empirical relation 
\( M_{\frac{1}{2} \rm dyn} = \beta{\sigma_{\rm e}}^2R_{\rm e}/G \) presented in 
\citet{Cappellari2006, Cappellari2013b}.  We use $r$-band photometry from the 
Siena Galaxy Atlas \citep{Moustakas2021}.  For galaxies in our sample, the Siena Galaxy 
Atlas has an average 5-$\sigma$ PSF detection depth in $r$-band of 26.19 AB mag, 
significantly deeper than the SDSS $r$-band photometry.  
We refer to the SDSS-$r-$band filter as the $r$ filter throughout the paper, 
and assume $r_{\rm SDSS}=r_{\rm DECam}+0.0381$~mag when calculating the 
luminosity based on Siena photometry. The mean offset between the 
luminosity from Siena and the SDSS {\tt cmodel} photometry is 
$\Delta \log(L_r/L_{\odot})$ = 0.06, suggesting that the luminosity from Siena is on 
average $14\%$ higher than SDSS.

In order to compare directly with the ATLAS$^{\rm 3D}$ sample, we adopt the relation given 
by \citet{Cappellari2006}, with $\beta=2.5$, in the following sections.  We note that there 
are three systematic effects that could cause a shift in $\beta$ for our sample.  First, as 
noted, our photometry is systematically deeper than the SDSS photometry used by
\citet{Cappellari2013}. 
Second, the methods in deriving $R_{\rm e}$ are different.
Third, our sample is dominated by slow rotators, unlike the vast
majority of the ATLAS$^{3D}$ sample \citep{Veale2017a}, and therefore may have systematically 
different dynamical masses for the same $\sigma$ and $R_{\rm e}$.
With $\beta=2.5$, the mean dynamical mass of galaxies in our sample is 
$\langle \log(M_{\rm dyn}/M_{\odot}) \rangle=11.66\pm0.18$.  The dynamical masses are shown 
in Column 4 of Table~1.  

Our goal is to compare the dynamical and stellar population-based masses, and for this 
we need an $M_{\star}/L$ within $R_{\rm e}$.  To estimate the luminosity-weighted $M_{\star}/L$ 
within $R_{\rm e}$, we first obtain $M_{\star}/L$ by modeling spectra extracted in effective 
circular apertures of $R_{\rm e}/8$,$R_{\rm e}/4$ and $R_{\rm e}/2$. Most galaxies in our sample have a 
declining $M_{\star}/L_r$ with increasing radii.  We will explore $M_{\star}/L$ 
gradients within individual galaxies in more detail in an upcoming paper. 
Since spectra from smaller aperture have higher S/N compared to the extraction in $R_{\rm e}$ 
and the high $S/N$ is important for the M/L measurement, for now the luminosity-weighted 
$M/L$ within $R_{\rm e}$ is derived from a linear extrapolation of $\log{R}-\log{(M_{\star}/L)_r}$ 
at $R_{\rm e}/8$,$R_{\rm e}/4$ and $R_{\rm e}/2$. The stellar masses are calculated using the luminosity 
from the Siena photometry where available or SDSS {\tt cmodel} (\S~2), and the results 
are shown in Table~1.  We discuss the comparison between dynamical mass and stellar mass 
within $R_{\rm e}$ in \S~5.

When we discuss the effective stellar surface mass density 
($\Sigma$) within $R_{\rm e}$,
\( \Sigma=M_{\star}/(2\pi R_{\rm e}^2)=(M_{\star}/L) \times L /(2\pi R_{\rm e}^2) \), 
we use the $(M/L)_{\rm Kroupa, R_{\rm e}}$, the mass-to-light ratio assuming a fixed 
Kroupa IMF, in order to discuss the relation between $\Sigma_{\rm Kroupa}$ and $\alpha_{\rm IMF}$. We do not use the best-fit $M/L$ within $R_{\rm e}$.
The reason is that since
\( \Sigma=M_{\star}/(2\pi R_{\rm e}^2)=(M/L) \times L /(2\pi R_{\rm e}^2)=\alpha_{IMF,R_{\rm e}}\times(M/L)_{\rm Kroupa} \times L /(2\pi R_{\rm e}^2) \), 
and the best-fit surface density from our fiducial model is actually 
$\Sigma=\alpha_{\rm IMF,R_{\rm e}} \times \Sigma_{\rm Kroupa}$, $\Sigma$ based on the best-fit $M/L$ will inevitably show a correlation with the extrapolated $\alpha_{\rm IMF}$ within $R_{\rm e}$, and bring any 
discussion about $\Sigma$ and $\alpha_{\rm IMF}$ into question.  Therefore we choose to 
use the estimated $M/L$ within $R_{\rm e}$ assuming a Kroupa IMF in order to exclude the 
effect of $\alpha_{\rm IMF, Re}$ when discussing the relation. We note that using a Kroupa IMF underestimates the 
stellar mass surface density and we will briefly discuss the differences brought with this assumption in later sections.  We also note the difference from the 
the stellar mass calculation (presented in Table~1, Table~2, and Figure~\ref{fig_mstar}) which uses the best-fit $M/L$.

% ---------------------------------------------------------------------------- %
\subsection{Central Properties of Massive Early-type Galaxies}

The 41 early-type galaxies in our sample are among the most massive galaxies in 
the local universe.  These galaxies have a mean central velocity dispersion of 
$\langle \sigma_{\rm c} \rangle=278\pm30$\kms \citep{Veale2017a}, an average velocity dispersion within $R_{\rm e}$ of $\langle \sigma_{\rm e} \rangle=255\pm28$\kms, and 
an average stellar mass of $\langle \log{(M_{\star}/M_{\odot})} \rangle=11.67\pm0.18$.  

We focus on the stellar population properties of individual 
galaxies within $R_{\rm e}/8$ throughout this work.  Table~1 lists the stellar populations 
and stellar IMF within $R_{\rm e}/8$ through detailed full spectral modeling.  
The error bars indicate the 16th and 84th percentiles of the posterior distributions.  
Galaxies in our sample cover a relatively narrow velocity dispersion 
($\langle \sigma \rangle=272\pm32$\kms), stellar and dynamical mass range.  Here we 
list the mean values of the whole sample for some properties that are not included 
in Table~1:  $\langle$[C/Fe]$\rangle=0.21\pm0.04$, $\langle$[N/Fe]$\rangle=0.17\pm0.09$, $\langle$[Ca/Fe]$\rangle=0.06\pm0.02$, $\langle$[Si/Fe]$\rangle=0.13\pm0.05$, $\langle$[Ti/Fe]$\rangle=0.15\pm0.06$. We also note that the best-fit 
[Na/Fe] is super-solar for all galaxies, with an average value of $\langle$[Na/Fe]$\rangle=0.44\pm0.10$.
We estimate the total metallicity using the equation 
from \citet{Thomas2003}:  ${\rm [Z/H]}={\rm [Fe/H]}+0.94\times{\rm [Mg/Fe]}$. The mean 
total metallicity over our sample is $\langle$[Z/H]$\rangle=0.38\pm0.04$.
Galaxies in our sample are dominated by the old stellar population with 
a mean young fraction of only $0.28\%$.

% newly added Figure~5
In Figure~\ref{fig_alpha-sig} we present $\alpha_{\rm IMF}$ as a function of $\sigma$ 
directly measured from the extracted spectra within $R_{\rm e}/8$.  We compare our results 
with $\alpha$ measurements from the recent literature using different methods: 
$\alpha_{\rm IMF}$ of A3D galaxies within $R_{\rm e}$ based on the best-fitting JAM model 
and NFW halo as a function of of 
central $\sigma$ ($R_{\rm e}/8$) \citep{Cappellari2013b}, central $\alpha_{\rm IMF}$ as a function 
of central $\sigma$ ($R_{\rm e}/8$) of 38 galaxies in \citep{Conroy2012b} based on full spectral 
modeling, results of 56 ETGs based on joint analysis of lensing and dynamical modeling 
\citep{Treu2010}, and stellar $M/L$ measured based on lensing and an assumption of dark 
matter fraction from EAGLE simulation in the SINFONI Nearby Elliptical Lens Locator 
Survey \citep{Newman2017}.  The reference IMF have been all converted to Kroupa.  
A general trend that $\alpha_{\rm IMF}$ increases with increasing $\sigma$ have been reported 
in \citet{Conroy2012b}, \citet{Cappellari2013b} and \citet{Treu2010}, and our results are in 
good agreement with the trend.

Figure~\ref{fig_sigmarel} shows the stellar population and IMF parameter fits as a function 
of $\sigma$. Error-bars indicate the 16th and 84th percentiles and points indicate the median 
of the posterior distribution.  Colors indicate the stellar mass. Since our sample 
contains only the most massive galaxies and covers a limited dynamic range, we 
complement our sample with low mass galaxies from CvD (see \S~2.2).  Their results are shown 
as green dots in Figure~\ref{fig_sigmarel}.  There is an apparent 
offset in [Fe/H] between our sample and CvD for galaxies in the regime of overlap with 
$\sigma\sim220$ to $260$\kms~.  
The model used in this work utilizes an updated stellar library \citep{Villaume2017} with 
wider metallicity coverage.  Therefore the discrepancy could be due to the difference in 
models. Moreover, since [Fe/H] has a well-known gradient with radius, differences in $R_{\rm e}$ 
measurements could add to the apparent discrepancy.  

In our sample, the average $M/L$ in the $r$ and $g$ bands are 
$\langle M/L_r \rangle=6.37\pm1.41 M_{\odot}/L_{\odot}$, and 
$\langle M/L_i \rangle=4.35\pm0.94 M_{\odot}/L_{\odot}$, respectively. We also calculate the 
$\alpha_{\rm IMF}$ mismatch parameter to indicate how much the IMF of our best-fit model deviates 
from our reference IMF. We take the Kroupa IMF as our reference, and find an average 
$\alpha_{\rm IMF}$ of $\langle \alpha_{\rm IMF} \rangle=\langle(M/L)/(ML)_{\rm MW} \rangle=1.84\pm0.43$.  
On average, the massive galaxies in our sample have an IMF that is more bottom 
heavy than one with a Salpeter slope, which has an $\alpha_{\rm IMF}=1.55$.  All of the galaxies 
have a stellar IMF that is more bottom heavy than Kroupa (Table 1). Galaxies in our sample 
span a large range of $\alpha_{\rm IMF}$ from $1.12^{+0.11}_{-0.10}$ (NGC~1700) 
to $3.05^{+0.39}_{-0.36}$ (NGC~4555).  The sample standard deviation is 0.43, 
much larger than the average measurement uncertainty of 0.26, therefore 
the scatter in $\alpha_{\rm IMF}$ is unlikely to be explained by 
measurement uncertainty alone.
We will discuss possible drivers for the scatter in the following sections.

% -------------------------------- %
% --Table2--
%\begin{center}
\begin{table*}[t]
\centering
\caption{Pearson Correlation Coefficients for MASSIVE data}
\renewcommand{\arraystretch}{1.5}
\begin{tabular}{@{} p{0.16\linewidth}|p{0.12\linewidth}p{0.12\linewidth}p{0.12\linewidth}p{0.12\linewidth}p{0.12\linewidth}p{0.12\linewidth} @{} }
\toprule
 & [Fe/H] & [Mg/Fe] & [Z/H] & $\log({\rm age})$ & $\log(M/L_r)$ & $\log(\alpha_{\rm IMF})$\\
 &   &   &  & [Gyr] & [$M_{\odot}/L_{\odot}$] &  \\
\hline
$\log(\sigma/{\rm km s}^{-1})$ & -0.13$\pm$0.17 & 0.44$\pm$0.11$^{**}$ & 0.22$\pm$0.13 & 0.38$\pm$0.14$^*$ & 0.40$\pm$0.13$^{**}$ & 0.08$\pm$0.16\\
$\log(\sigma_c/{\rm km s}^{-1})$ & -0.09$\pm$0.18 & 0.39$\pm$0.12$^*$ & 0.22$\pm$0.14 & 0.32$\pm$0.15$^*$ & 0.37$\pm$0.16$^*$ & 0.14$\pm$0.17\\
$\log(\sigma_e/{\rm km s}^{-1})$ & 0.12$\pm$0.22 & 0.22$\pm$0.19 & 0.31$\pm$0.12$^*$ & 0.03$\pm$0.25 & 0.25$\pm$0.14 & 0.18$\pm$0.19\\
$\log(M_{\star}/M_{\odot})$ & -0.30$\pm$0.14 & -0.04$\pm$0.14 & -0.35$\pm$0.12$^*$ & 0.39$\pm$0.17$^{**}$ & 0.35$\pm$0.19$^*$ & 0.17$\pm$0.17\\
$\log(M_{dyn}/M_{\odot})$ & -0.34$\pm$0.15$^*$ & 0.03$\pm$0.17 & -0.33$\pm$0.14$^*$ & 0.35$\pm$0.14$^*$ & -0.05$\pm$0.23 & -0.20$\pm$0.17\\
$\log(M_{h}/M_{\odot})$ & -0.41$\pm$0.16$^*$ & 0.25$\pm$0.17 & -0.24$\pm$0.17 & 0.34$\pm$0.19$^*$ & 0.12$\pm$0.25 & -0.02$\pm$0.22\\
$\log(L_r/L_{\odot})$ & -0.10$\pm$0.15 & -0.23$\pm$0.12 & -0.29$\pm$0.14 & 0.03$\pm$0.13 & -0.21$\pm$0.14 & -0.21$\pm$0.15\\
$\log(\nu_{10})$ & -0.36$\pm$0.12$^*$ & 0.23$\pm$0.15 & -0.20$\pm$0.14 & 0.24$\pm$0.11 & -0.01$\pm$0.15 & -0.16$\pm$0.15\\
$\log(1+\delta_g)$ & -0.21$\pm$0.15 & 0.17$\pm$0.16 & -0.09$\pm$0.17 & 0.05$\pm$0.12 & -0.10$\pm$0.13 & -0.14$\pm$0.15\\
$\log(\Sigma_{\rm Kroupa}/(M_{\odot}{\rm kpc}^{-2}))$ & 0.34$\pm$0.17$^{\star}$ & 0.12$\pm$0.17 & 0.46$\pm$0.12$^{**}$ & -0.17$\pm$0.13 & 0.28$\pm$0.16 & 0.34$\pm$0.14$^*$\\
$\log(R_e/{\rm kpc})$ & -0.37$\pm$0.16$^*$ & -0.08$\pm$0.18 & -0.47$\pm$0.12$^{**}$ & 0.27$\pm$0.13 & -0.15$\pm$0.20 & -0.25$\pm$0.16\\
\hline
\end{tabular}
\begin{tablenotes}
\item $^{**}$ Correlation is significant at the 0.01 level
\item $^{*}$ Correlation is significant at the 0.05 level
\end{tablenotes}
\vspace{0.1cm}
\end{table*}
%\end{center}
% -------------------------------- $$

\subsection{Stellar Population Scaling Relations in the MASSIVE Sample}

In Figure~\ref{fig_sigmarel} and Table~2 we present the stellar population 
and IMF parameters as a function of $\sigma$. We calculate Pearson 
correlation coefficients ($r$) between $\sigma$ and all the stellar population 
or IMF parameters.  We calculate the $p$-value for each 
correlation, which is the probability that the correlation is 
produced by an uncorrelated system.  The parameter pairs with a $p$-value smaller 
than $5\%$ are shown in black in the corresponding panels, where the 
uncertainty comes from 1000 bootstrap samples.  The correlations measured 
based on the combined sample are shown in green.

Table~2 shows the correlation between two sets of variables.  The first 
set includes stellar populations and the stellar IMF. The second set includes 
$\sigma$, stellar mass and dynamical mass, luminosity, and effective stellar 
surface density, and others.  Based on the $p-$values, significant correlations are 
marked with an asterisk.  For galaxies in our sample, we see mild positive 
correlations between $\log(\sigma)$ and the following parameters: [Mg/Fe], 
[O/Fe], [C/Fe] and [Na/Fe], as shown in Figure~\ref{fig_sigmarel}. 

The trend that metallicity and [Mg/Fe] in ETG centers increase 
with increasing galaxy velocity dispersion has been shown in previous 
literature \citep[e.g.,][]{Trager2000b, Worthey2003, Thomas2005, Conroy2014}. 
Within our sample there is no 
strong positive correlation between $\log(\sigma)$ and [Fe/H]. This could 
be due to the intrinsic scatter in [Fe/H] at the high mass end and our 
limited dynamical range.  In addition, aperture effects 
could also wash out trends between $\sigma$ and [Fe/H] since there is a 
strong negative [Fe/H] gradient within galaxies. [Fe/H] could be sensitive 
to the choice of aperture and the measurement of $R_{\rm e}$.

In addition, we note that the correlations between $\sigma$ and the following parameters 
are not included in the table or figure: [C/Fe]--$\log(\sigma)$: $r=0.49\pm0.12$; 
[N/Fe]--$\log(\sigma)$: $r=0.27\pm0.15$; [Si/Fe]--$\log(\sigma)$: $r=0.12\pm0.17$. 
Despite the limited dynamic range, we found that [Mg/Fe], [C/Fe], [Ca/Fe] 
all increase within increasing $\sigma$ and they are consistent with previous studies 
\citep[e.g.][]{Graves2007, Thomas2010, Johansson2012b, Greene2015, Conroy2014}. 
The trends of increasing abundances with increasing $\sigma$ for O, Na 
are consistent with \citet{Conroy2014}.  [C/H] and [O/H] are strongly correlated 
in our sample with $r\approx0.87$.

Although all galaxies in our sample have an IMF that is bottom heavier than Kroupa, 
there is a large scatter in $\alpha_{\rm IMF}$, and basically no correlation between 
$\log(\sigma)$ and $\alpha_{\rm IMF}$. Some of the correlations involving the IMF are 
shown in Figure~\ref{fig_mstar}. Within our sample, the effective 
stellar mass surface density ($\log(\Sigma_{\rm Kroupa})$) (Table~2) and 
the total metallicity ([Z/H], Table~3) appear to positively correlate 
with $\log(\alpha_{\rm IMF})$ 
with $r=0.34\pm0.14$ and $r=0.25\pm0.17$, respectively. There are  
moderate correlations in $\log(\Sigma)$--[Fe/H] and $\log(\Sigma)$--[Z/H] of 
$r=0.34\pm0.17$ and $r=0.46\pm0.12$, respectively, 
indicating that in our sample, more compact galaxies are more metal rich and 
have a bottom-heavier IMF.  We do not see any significant 
correlations between $\log(\alpha_{\rm IMF})$ and [Mg/Fe], [O/Fe], [Ca/Fe] or [Ti/Fe] 
within our sample.  The correlation with [Na/Fe] is moderate (Table~3). 
We note that when calculating $\Sigma_{\rm Kroupa}$ we adopt the assumption 
of a Kroupa IMF, which underestimates the stellar mass (\S~4.1). When we use 
the stellar mass calculated in our fiducial model (adopting an $M/L$ 
extrapolated to $R_{\rm e}$), the $\log(\Sigma)$--$\log(\alpha_{\rm IMF})$ 
correlation becomes more significant, with $r=0.47\pm0.10$, $p=0.4\%$.  
There is suggestive evidence that galaxy compactness 
might be an important property related to IMF variation. 

% -------------------------------- %
% --Table3 -- %
\begin{table}[b]
\caption{}
\renewcommand{\arraystretch}{1.5}
%\centering
\resizebox{0.98\width}{!}{\begin{tabular}{l|cc|r}
\toprule
MASSIVE  & $\log(M/L_r)$ & $\log(\alpha_{\rm IMF})$ & $\log(\alpha_{\rm IMF})$\\
\hline
[Fe/H] & -0.17$\pm$0.21 & 0.24$\pm$0.17& 0.26$\pm$0.15\\

[Mg/Fe] & 0.32$\pm$0.14$^*$ & 0.00$\pm$0.16& -0.05$\pm$0.16\\

$\log(\sigma/{\rm km s}^{-1})$ & 0.41$\pm$0.13$^{**}$ & 0.08$\pm$0.16 & $-$\\ 

[Z/H] & 0.08$\pm$0.20 & 0.25$\pm$0.17& 0.25$\pm$0.15\\

[O/Fe] & 0.19$\pm$0.12 & -0.05$\pm$0.15& -0.09$\pm$0.16\\

[Na/Fe] & 0.36$\pm$0.12$^*$ & 0.25$\pm$0.16& 0.23$\pm$0.15\\

[Ca/Fe] & 0.19$\pm$0.14 & 0.01$\pm$0.13& -0.03$\pm$0.16\\

[Ti/Fe] & -0.12$\pm$0.16 & -0.19$\pm$0.15& -0.21$\pm$0.15\\

$\log(\Sigma_{\rm Kroupa}/(M_{\odot}{\rm kpc}^{-2}))$ & 0.29$\pm$0.16 & 0.34$\pm$0.15$^*$& 0.36$\pm$0.15$^*$\\

\toprule
MASSIVE+CvD   & $\log(M/L_r)$ & $\log(\alpha_{\rm IMF})$ & $\log(\alpha_{\rm IMF})$\\
\hline
[Fe/H] & 0.18$\pm$0.11 & 0.31$\pm$0.10$^{**}$& 0.15$\pm$0.11\\

[Mg/Fe] & 0.73$\pm$0.06$^{**}$ & 0.66$\pm$0.08$^{**}$& 0.44$\pm$0.09$^{**}$\\

$\log(\sigma/{\rm km s}^{-1})$ & 0.69$\pm$0.06$^{**}$ & 0.56$\pm$0.07$^{**}$ & $-$\\ 

[Z/H] & 0.49$\pm$0.09$^{**}$ & 0.56$\pm$0.08$^{**}$& 0.32$\pm$0.10$^{**}$\\

\hline
\end{tabular}

}
%\begin{tablenotes}
%\end{tablenotes}
\tablecomments{Pearson Correlation Coefficients (left) and Partial Correlation 
Coefficients, controlling $\log(\sigma)$ (right) for MASSIVE data (top), 
and the combined sample MASSIVE+CvD (bottom)\\
$^{**}$ Correlation is significant at the 0.01 level\\
$^{*}$ Correlation is significant at the 0.05 level\\}
%\vspace{0.1cm}
\end{table}
% -------------------------------- %

The galaxies with the most bottom-heavy IMF in Figure~\ref{fig_mstar} have moderate 
stellar mass and luminosity, but typically all have high effective stellar mass surface 
density, i.e., their distinguishing feature is their compactness instead of how massive 
they are. We will study how compactness affect the local or global $\alpha_{\rm IMF}$ in 
detail in an upcoming paper. The effective surface density, $\Sigma$, requires a 
measurement of both $R_{\rm e}$ and $L$, which cannot be done uniformly across the two samples. 
Therefore our analysis of trends with $\Sigma$ is limited to the MASSIVE sample.  
Since there is an overall trend that the IMF becomes bottom heavier in more massive 
galaxies (\S~4.4),  we are also interested in whether the correlation with IMF 
within our sample is driven by $\log(\sigma)$. The partial correlation within the MASSIVE 
sample in $\log(\Sigma)$--$\log(\alpha_{\rm IMF})$ 
is $r=0.36$,$p=0.04$ when holding $\log(\sigma)$ and [Mg/Fe] constant, 
indicating the connection between $\log(\Sigma)$ and $\log(\alpha_{\rm IMF})$ 
is significant and independent of $\log(\sigma)$ and [Mg/Fe].
%Both [Z/H] and $\log(\Sigma)$ are moderately correlated with $\log(\sigma)$, 
In the rightmost column of Table~3 we present the partial correlation coefficients by 
fixing the effect of $\log(\sigma)$, which is $r\approx0.25$ for [Z/H]-$\log(\alpha_{\rm IMF})$ and 
$r\approx0.36$ for $\log(\Sigma)$-$\log(\alpha_{\rm IMF})$, suggesting 
these positive correlations are not driven by galaxy central velocity dispersion.

In summary, within our sample, there is no significant correlation between 
$\alpha_{\rm IMF}$ and $\sigma$ or [Mg/Fe].  Both [Z/H] and $\log(\Sigma)$ 
show moderate positive correlations with $\alpha_{\rm IMF}$, and $\log(\Sigma)$ is
moderately correlated with [Z/H], suggesting that they may be both responsible 
to the IMF variation within our sample

\subsection{Stellar Population Scaling Relations in the Combined Sample}

We present the correlations in the combined sample in Figure~\ref{fig_alpharel} 
and Table~3. In Figure~\ref{fig_alpharel}, colors indicate the effective stellar 
mass surface density. Pearson correlation coefficients are shown in green at the top right 
corner of each panel for the combined sample. By including galaxies with lower $\sigma$ from CvD, 
we find moderate positive correlations between $\log(\alpha_{\rm IMF})$ and $\log(\sigma)$, 
[Mg/Fe], and [Z/H]. Among them, [Mg/Fe] has the most significant correlation with the
IMF, while the relation between $\log(\alpha_{\rm IMF})$ and [Fe/H] or $\log(\sigma)$ is mild.
However, we recall from \S~2.2 that there is some suspicious differences in 
[Fe/H] between the two samples.  Until we have a sample covering the full 
range in $\sigma$ and analyzed in a uniform manner, we  note this caveat 
in the [Fe/H] and [Z/H] correlations with $\alpha_{\rm IMF}$.

To study the correlation between $\alpha_{\rm IMF}$ and $\log(M_{\star})$, 
$\log(L_r)$ and $\log(M_{\rm dyn})$, we further combine our sample 
with the galaxies that overlap between CvD and \citet{Cappellari2013}. 
Specifically, we make use of the stellar mass, dynamical mass, 
and luminosity measurements in \citet{Cappellari2013}. 
In the combined sample, $\log(M_{\star})$, $\log(L_r)$ and $\log(M_{\rm dyn})$ 
all have moderate positive correlations with $\log(\alpha_{\rm IMF})$.  The 
correlation coefficients are $r=0.50\pm0.08$, $0.40\pm0.09$ and $0.42\pm0.09$, 
respectively. Over the wide dynamic range afforded by the combined sample, 
we confirm that in general the IMF in more massive galaxies is more bottom heavy.

We next address whether the correlation between the IMF parameters and metallicity 
or [Mg/Fe] are driven by galaxy central velocity dispersion. The partial correlations 
at fixed $\log(\sigma)$ are shown in the the rightmost column of Table~3, suggesting 
that both [Z/H] and [Mg/Fe] are moderately correlated with $\log(\alpha_{\rm IMF})$ when 
the effect of $\sigma$ is fixed.  Therefore, although the correlation between galaxy 
central velocity dispersion and stellar population plays a role here, these 
moderate correlations with [Mg/Fe] and [Z/H] suggest that [Mg/Fe] and total metallicity 
are both driving IMF variation in a way that is independent of the effect from galaxy 
central velocity dispersion. 

In summary, by complementing our sample with lower masses galaxies in CvD and 
\citep{Cappellari2013}, we find that the IMF of ETGs becomes increasingly bottom heavy 
with increasing central velocity dispersion, luminosity, stellar mass, and dynamical mass. 
[Mg/Fe] and [Z/H] are positively correlated with the IMF and 
these trends are not driven by their relation to the galaxy central velocity dispersion.

% ---------------------------------------------------------------------------- %
% moved from section 6.1
% --------------------------------
\begin{table}[h]
\caption{Best-fit parameters}
\renewcommand{\arraystretch}{1.5}
\centering
\resizebox{1.05\width}{!}{% results from bi-variate fitting
\begin{tabular}{l|ccc}
\toprule
Eq (1) & $c_0$ & $c_1$ & $c_2$\\
\hline
MASSIVE & $0.-1\pm0.87$ & $0.10\pm0.38$ & $-0.04\pm0.46$ \\ 
CvD & $-0.36\pm0.34$ & $0.03\pm0.16$ & $1.91\pm0.33$ \\ 
MASSIVE+CvD & $-0.66\pm0.34$ & $0.24\pm0.16$ & $1.10\pm0.26$ \\ 
\toprule
Eq (2) & $c_0$ & $c_1$ & $c_2$\\
\hline
MASSIVE & $0.07\pm0.78$ & $-0.01\pm0.33$ & $0.69\pm0.37$ \\ 
CvD & $-1.25\pm0.43$ & $0.57\pm0.19$ & $0.45\pm0.43$ \\ 
MASSIVE+CvD & $-1.02\pm0.34$ & $0.47\pm0.15$ & $0.37\pm0.13$ \\ 
\hline
\end{tabular}}
\begin{tablenotes}
\item {Eq (1): \( \log(\alpha_{\rm IMF})=c_0+c_1(\log\sigma)+c_2({\rm [Mg/Fe]}) \)} 
\item {Eq (2): \( \log(\alpha_{\rm IMF})=c_0+c_1(\log\sigma)+c_2({\rm [Z/H]}) \)}
\end{tablenotes}
\end{table}
% --------------------------------
\subsection{Multivariate Linear Regression}

We have presented the correlations between stellar populations, IMF parameters, 
and galaxy dynamical properties.  We find that the effective velocity dispersion 
($\sigma$), [Mg/Fe], [Z/H] and possibly the effective stellar mass surface 
density ($\Sigma_{\rm Kroupa}$),  are all correlated with $\alpha_{\rm IMF}$.  Now we 
examine the relative strength of these correlations by fixing the effects of certain 
variables.  Following the analysis by \citet{Smith2014}, we first focus on $\sigma$ 
and [Mg/Fe], since they are shown to be strongly correlated with $\alpha_{\rm IMF}$ 
in the combined sample.  We perform a bi-variate linear regression of $\log(\sigma)$, 
[Mg/Fe] and $\log(\alpha_{\rm IMF})$ in three datasets: our sample alone, the CvD alone, 
and the combined sample. Variables are standardized prior to the fitting.  
In the equation 
\begin{equation} 
\log(\alpha_{\rm IMF})=c_0+c_1(\log\sigma)+c_2({\rm [Mg/Fe]})
\end{equation}
The best-fit parameters fits to all three samples are shown in Table~4.

Based on the combined sample, the best-fit parameters suggest that 
[Mg/Fe] plays a more important role in driving the variation of
$\alpha_{\rm IMF}$ than $\sigma$.  The dominant role of [Mg/Fe] using a 
bi-variate model of the combined sample is consistent with CvD and \citet{Smith2014}. 
It suggests that [Mg/Fe] is the more important driver for 
$\log(\alpha_{\rm IMF})$ when considering both galaxy central velocity dispersion 
and [Mg/Fe] based on stellar population synthesis.  However within our 
sample, these two variables alone hardly describe the variance of 
$\log(\alpha_{\rm IMF})$, indicating 
that other variables must also be considered. 

Based on the strong correlation between IMF and metallicity 
within our sample and in the combined sample, we further test the bi-variate 
linear regression with $\log(\alpha_{\rm IMF})$ as the dependent variable using 
$\log(\sigma)$ and [Z/H] as independent variables using the following equation  
\begin{equation} 
\log(\alpha_{\rm IMF})=c_0+c_1(\log\sigma)+c_2({\rm [Z/H]})
\end{equation}
The results from the combined sample suggest that total metallicity and 
$\log(\sigma)$ are of similar importance in driving the IMF variation. We will discuss the physical implications in \S~6. 
As discussed in \S~4.3, $\Sigma$ is also an important property for the IMF and 
may be a driver of IMF variations at the high mass end.  However, in this work 
we do not include $\Sigma$ in the discussion since it is sensitive to 
the consistency in the measurements of luminosity and $R_{\rm e}$. 

% ---------------------------------------------------------------------------- %
\begin{figure*}[t]
  \centerline{\psfig{file=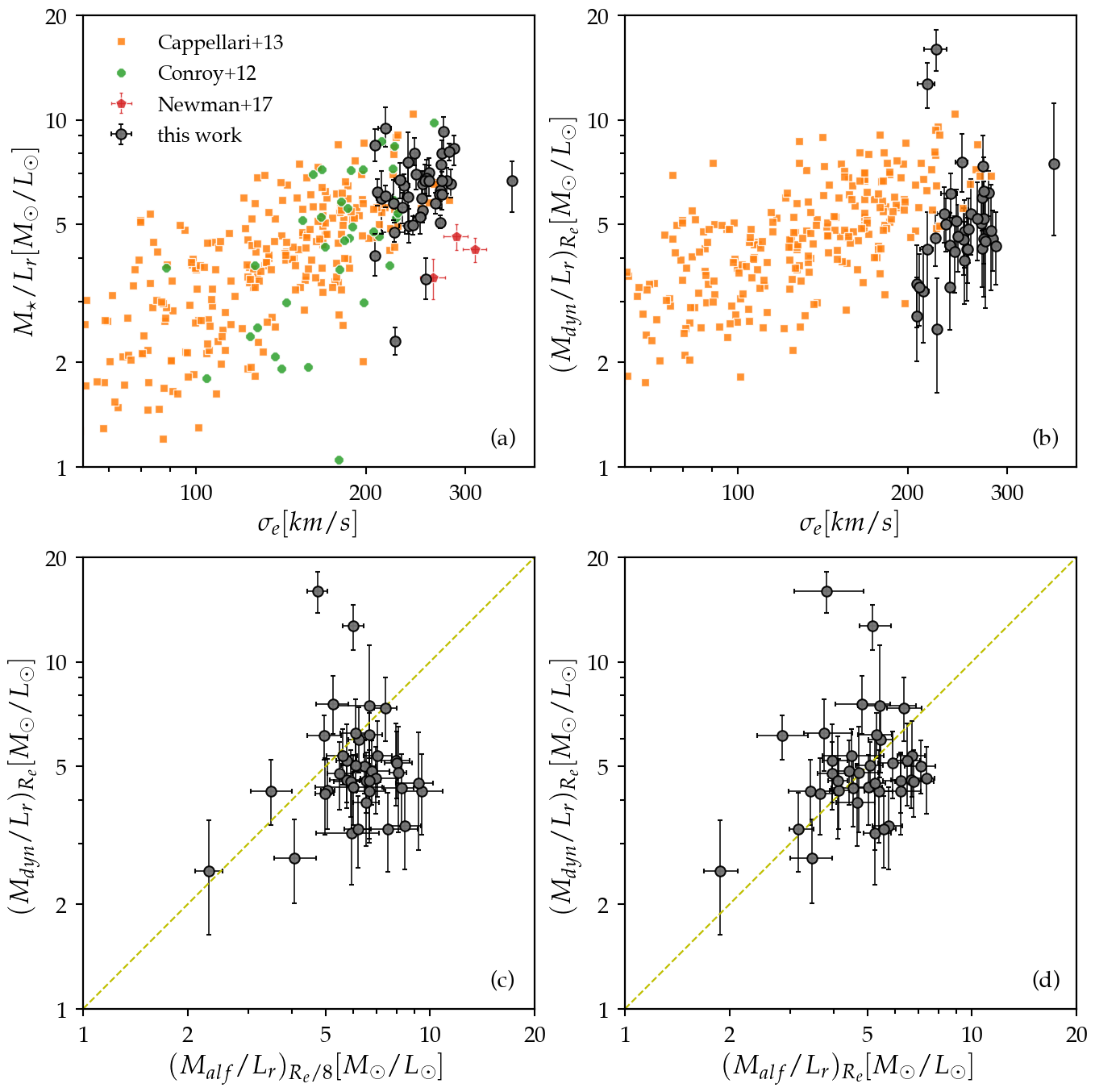, 
  width=15cm}}
  \caption{ 
  Panel (a):  $r-$band stellar $M/L$ in galaxy centers as 
  a function of $\sigma_{\rm e}$. Results from this work (black) are compared to the 
  estimated stellar population M/L in A3D (orange dots), 
  in \citet{Newman2017} (blue) and in CvD (green).  
  Panel (b): Dynamical $M/L$ in $r$-band as a function of $\sigma_{\rm e}$. 
  Panel (c):  Comparison between the $M/L$ from {\tt alf} and the estimated 
  dynamical$M/L$M/L.  Note that the stellar M/L are derived in the aperture of 
  $R_{\rm e}/8$, and the dynamical $M/L$ indicate an average value within $R_{\rm e}$.  
  Yellow dashed and dotted lines indicates 1:1 comparisons.  
  Panel (d): $M_{alf}/L_r$ are extrapolated to an aperture of $R_{\rm e}$.  
  The majority galaxies in our sample are consistent with $M_{\rm dyn}/L_r$ estimation 
  within 1$\sigma$ uncertainty.
  }
\label{fig_m2l}
\end{figure*}
% ---------------------------------------------------------------------------- %
\subsection{Environment}

In Table~2 we also present [Fe/H], [Mg/Fe] and the IMF $\alpha-$mismatch parameter 
as a function of three environment indicators: logarithmic halo mass ($M_h$), 
local over-density ($\nu_{10}$) and large-scale galaxy density($1+\delta_g$) 
\citep{Veale2017a}. $\nu_{10}$ represents the luminosity density 
of galaxies in a sphere enclosing the $10^{\rm th}$ nearest neighbour, and 
 $\delta_g$ represents the luminosity-weighted galaxy density contrast with a 
 smoothing scale of 5.7~Mpc. 

There is no significant correlation between $\alpha_{\rm IMF}$ and 
any environmental indicator within our sample.  This is consistent with what was found in \citet{Rosani2018}.
However, we do find a negative correlation between [Fe/H] and the logarithmic 
dynamical mass within our sample, which is not consistent with the significant positive 
correlation found in the combined sample where $r_{\rm[Fe/H]-\log(M_dyn)} = 0.48, p = 2.7\times10^{-5}$, i.e., over a large dynamic range the massive ETGs in our sample 
are still found to be more metal rich 
compared to lower mass ETGs. We note that the negative correlation within our sample  
may be at least partially due to differences in $R_{\rm e}$ measurements, as 
[Fe/H] is sensitive to the choices of apertures.  In an upcoming paper, we 
will study the local variations in stellar population parameters, the IMF, and their relations 
with environmental indicator. 

With the large dynamical range in the combined sample, [Mg/Fe] seems to be 
an important parameter for the IMF.  In upcoming work, we will expand our 
investigation to larger radii where galaxy properties are dominated 
by accretion and could be more sensitive to the environment.  We will study the 
relation between local IMF or IMF gradient with [Mg/Fe] and explore any connection with the environment.

% ---------------------------------------------------------------------------- %
\section{Dynamical Versus Stellar Masses}

While independent modeling techniques from stellar population synthesis and dynamical 
modeling both reveal similar global IMF trends. The 
concerning aspect arises when comparing the IMF constraints from different methods, because
the inferred $M/L$ are not consistent on a galaxy-by galaxy basis 
\citep[e.g.][]{Smith2014, Newman2017}. \citet{Smith2014} compare the stellar $M/L$ in the 
overlap sample between \citet{Conroy2012b} and \citet{Cappellari2013}, and conclude that 
there is no significant correlation between the stellar $M/L$ or IMF mismatch parameter 
inferred by the two studies.  \citet{Smith2014} mention that $M/L$ and $\alpha_{\rm IMF}$ 
variations within galaxies could lead to discrepancies due to the different apertures used 
for the dynamical and stellar population synthesis studies.

At the very least, the stellar-population--based masses should not violate the mass budget. 
Here we compare the stellar $M/L$ with the estimated total $M_{\rm dyn}/L_r$ to ensure that our 
inferred stellar masses are physical.  As described in \S~4, we estimate the stellar $M/L$ 
within $R_{\rm e}$ with a linear extrapolation of $\log{R}-\log{(M/L)_r}$. 
Since the dynamical measurements are done within $R_{\rm e}$, it is important to account for the $M/L$ variation within galaxies. 
Most galaxies in our sample have a declining radial profile of $M/L$, and on 
average the ratio between central $M/L$ and extrapolated $M/L$ within $R_{\rm e}$ is 1.3.
We compare our results with recent literature in Figure~\ref{fig_m2l} and present the main 
result in this section in Figure~\ref{fig_m2l}(d). 
 
In Figure~\ref{fig_m2l}(a), we compare the stellar $M/L$ with several measurements in the 
literature.  It shows the stellar $M/L_r$ in galaxy centers as a function of luminosity 
weighted velocity dispersion within $R_{\rm e}$.  Black data points show our results where 
the $M/L$ is measured within $R_{\rm e}/8$.  Orange data points show the $M/L_r$ of the stellar 
components within a sphere of radius $r\sim R_{\rm e}$ from \citet{Cappellari2013}, where the results 
were from the best-fitting JAM model with an assumption of NFW halo for the 
dark matter components. Green data points show the $M/L_r$ from full spectral modeling within a 
radius of $R_{\rm e}/8$ from CvD.  Blue data points show results from \citet{Newman2017}. 
The $M/L$ of the three galaxies are constrained by lensing and within an aperture of 
2.2\asec~ (1.4-2.2kpc), with the dark matter contribution estimated from the EAGLE simulations. 
Note that unlike all other data points in the sample, the velocity dispersions of these 
three galaxies are measured in an aperture of $R_{\rm e}/2$. There seems to be an offset 
between our results and the three galaxies from \citet{Newman2017}. 
The apertures correspond to 0.2 to 0.7 $\times R_{\rm e}$ and are all 
larger than $Re/8$ in our sample.  If $\alpha_{\rm IMF}$ declines with radius, then putting 
the galaxies in our sample with $R_{\rm e}/8$ apertures on the same scale as the 
three galaxies from \citet{Newman2017} may lower the $\alpha_{\rm IMF}$, and could possibly 
make the two sets of data in better agreement.

Figure~\ref{fig_m2l}(b) shows the dynamical $M/L_r$ as a function of $\sigma_{\rm e}$. Our results 
are compared with \citet{Cappellari2013} (orange data points) whose measurements come from 
the best-fitting self-consistent JAM model.  As described in \S~4, we use an empirical relation 
from \citet{Cappellari2006} to estimate $M_{\rm dyn}$. The correlation is confirmed by 
\citet{Cappellari2013}.  There is an apparent offset in the overlapping $\sigma_{\rm e}$ region. 
There are likely many factors contributing to this apparent offset. Our photometry is based on 
deeper imaging, and our $R_{\rm e}$ measurements may be systematically different as well. Furthermore, 
our sample is dominated by slow rotators, which perhaps requires a different virial factor $\beta$.
With the empirical relation from \citet{Cappellari2006}, Figure~\ref{fig_m2l}(b) suggests 
that we may systematically underestimate the dynamical mass of 
our galaxies compared to galaxies with similar $\sigma_{\rm e}$ in ATLAS$^{\rm 3D}$. 
We consider several possibilities that could mitigate the difference: 
First, the luminosity and structural measurements  
both matter. For the galaxies with both SDSS and Siena photometry, if we use the size 
and luminosity from SDSS {\tt cmodel}, the $M_{\rm dyn}/L$ would increase by $\sim14\%$, 
suggesting that both the $R_{\rm e}$ measurement and the depth affect the $M_{\rm dyn}/L$ measurement. 
Considering that galaxies in our sample have 
extended stellar structures, deep photometry is crucial to determine the sky background 
level and light profiles, therefore we choose to adopt the Siena photometry described in 
\S~2 where available.  Second, the mean dynamical $M/L_r$ in our sample is 
${\langle M_{\rm dyn}/L_r \rangle} = 5.3\pm2.4 M_{\odot}/L_{r,\odot}$.  
If we use $\beta=4$ from \citet{Wolf2010}, the mean dynamical $M/L_r$ will increase by $60\%$.
Currently we do not have more accurate dynamical mass measurements and do not have better 
galaxy-by-galaxy constraints on $\beta$, therefore we adopt $\beta=2.5$ 
from A3D. In the future we will work on an updated calibration of the relation for 
virial mass estimation based on detailed dynamical modeling of ETGs in our sample. 
New results from Jeans Anisotropic models \citep{Cappellari2008} or 
Schwarzschild orbit superposition models \citep{Schwarzschild1979,Thomas2016} may 
help alleviate the discrepancy. 

From Figure~\ref{fig_m2l}(a), our results of $M_{\star}/L$ are consistent with CvD and A3D 
in the regime of $\sigma_{\rm e}$ from 220 to 300 \kms~ and indicate galaxies in this regime are 
better described by a bottom heavy IMF.  In Figure~\ref{fig_m2l}(c) we compare the dynamical 
mass-to-light ratio ($M_{\rm dyn}/L$) within $R_{\rm e}$ with $M_{\star}/L$ measured directly within 
$R_{\rm e}/8$: Most galaxies (31 out of 39) in our sample have a dynamical $M/L_r$ 
that is smaller than the stellar $M/L_r$ within $R_{\rm e}/8$, therefore it seems there's an apparent 
disagreement and violation of mass budget if we do not use consistent apertures.  However, this 
is partly due to the variation of $M/L$ within galaxies.  In Figure~\ref{fig_m2l}(d), we further 
compare the dynamical $M/L_r$ within $R_{\rm e}$ to the extrapolated stellar $M/L_r$ (\S~4.1) within $R_{\rm e}$. 
The mean $M_{\rm dyn}/L$ and $M_{\star}/L$ within $R_{\rm e}$ are $5.3\pm2.4 \, M_{\odot}/L_{\odot}$ 
and $5.0\pm1.2 \, M_{\odot}/L_{\odot}$, respectively, and are both smaller compared to 
the mean central $M_{\star}/L$ of $6.4\pm1.4 \, M_{\odot}/L_{\odot}$.  
If we use a consistent aperture of $R_{\rm e}$, most galaxies in our sample have $M_{\star}/L_r$ 
smaller than the $M_{\rm dyn}/L_r$ (20 out of 39), and most galaxies have consistent 
$M_{\star}/L_r$ and $M_{\rm dyn}/L_r$ within 1-$\sigma$ uncertainty (30 out of 39).  
Our result suggests that overall our stellar $M/L_r$ constraints are within the mass budget.  
Figure~\ref{fig_m2l} highlights the importance of using consistent apertures when comparing 
different works.  

% ---------------------------------------------------------------------------- %
\section{Discussion}

We have presented the correlations between stellar populations, IMF and galaxy 
dynamical properties in \S~4.  In \S~6.1 we discuss the physical implication.  
In \S~6.2 we compare our fiducial model with the 2pl IMF model where the low 
cutoff is fixed to $0.08M_{\odot}$.

\subsection{Physical Implications}

By combining our sample with low mass galaxies in \S~4.4 we present positive 
correlations between $\log(\alpha_{\rm IMF})$ and $\sigma$, $\log(M_{\star})$, 
$\log(L_r)$ and $\log(M_{\rm dyn})$, and confirm the trend towards an increasing 
$\alpha_{\rm IMF}$in more massive systems found in prior work 
\citep[e.g.][]{Conroy2012b, Cappellari2013, Spiniello2011, Treu2010, LaBarbera2013}, 
i.e., the IMF in ETG centers becomes increasingly bottom heavy with increasing 
galaxy masses.  
In addition, we have found that $\log(\alpha_{\rm IMF})$ is moderately correlated with 
$\log(\Sigma_{\rm Kroupa})$ within our sample (\S~4.3), and also moderately correlated 
with [Mg/Fe] and total metallicity [Z/H]in the combined sample (\S~4.4).  In this section,
we compare our results with previous literature and discuss the physical implications.

There has been a lot of debate on the physical drivers of IMF variation. 
Many recent observations reveal a correlation between IMF and galaxy stellar metallicity
\citep[e.g.][]{Martin-Navarro2015, Zhou2019, vanDokkum2017, Parikh2018}.  On the other hand, 
CvD reported a correlation between $\alpha_{\rm IMF}$ and both $\sigma$ and [Mg/Fe], but only 
a mild correlation with total metallicity.  Also some dynamical analysis suggest no 
significant correlation with metallicity or [Mg/Fe] \citep[e.g.][]{McDermid2014, Li2017}. 
There is not yet a clear conclusion on this topic.
One difference between our results and many previous studies is that we find clear 
evidence that {\it both} the central metallicity and [Mg/Fe] are positively correlated 
with the IMF.  In the combined sample, among all the parameters, [Mg/Fe] is the one that 
has the strongest correlation with $\alpha_{\rm IMF}$.  As shown in Figure~\ref{fig_alpharel}, 
the [Mg/Fe] in our sample is on average higher than the lower mass galaxies in CvD.  
If we use [Mg/Fe] to trace the $\alpha$-abundance and use it as an indicator of the past star 
formation timescale \citep[e.g.][]{Thomas2005}, the average 
$\langle {\rm [Mg/Fe]} \rangle = 0.29\pm0.04$ suggests that in the centers ($R_{\rm e}/8$) of the massive 
ETGs in our sample, the average star formation timescale is only $\Delta{t}\sim250$~Myr, 
inferring an average star formation rate surface density ($M_{\star}/(4\pi R^2 \Delta{t})$) 
of $117 M_{\odot}{\rm yr}^{-1}{\rm kpc}^{-2}$.  Based on these estimates, the stars in the 
central region of these ETGs are formed in extreme environments through a starburst 
\citep[e.g.][]{Kennicutt2012, Bouche2007, Daddi2010}.

The $\log(\Sigma)-{\rm [Fe/H]}$ and $\log(\Sigma)-{\rm [Z/H]}$ correlations 
(\S~4.3, Table~2) within our sample are also important, as both metallicity and stellar 
surface density are positively correlated with $\log(\alpha_{\rm IMF})$.  Within our sample, 
the metallicity has a significant positive correlation with the stellar surface density 
and a mild negative correlation with logarithmic stellar mass (Table~2). Our finding 
is consistent with the picture that higher density ETGs retain more of their metals 
and are less likely to be disrupted \citep[e.g.][]{Barone2018}.
The connection between the IMF and galaxy stellar metallicity has been shown in prior work 
both from observations \citep[e.g.][]{Martin-Navarro2015, Zhou2019, vanDokkum2017, Parikh2018}, 
and simulations \citep[e.g.][]{Sharda2021, Chon2021}. For example, \citet{Sharda2021} studied the 
characteristic mass in collapsing dusty gas clouds and found that high pressure 
ISM and high metallicity will result in low characteristic mass and therefore explained the 
bottom-heavy IMF found in massive ETGs.  Since the high surface density suggests high gas pressure 
when the stars are formed, the simulation result is supported by our data and explains that 
the metallicity and stellar surface density are driving the IMF variation in different ways.

The connection between galaxy compactness and the stellar IMF has also been 
suggested in several recent papers \citep[e.g.][]{Chabrier2014, LaBarbera2019, 
Smith2015, Barbosa2021, vanDokkum2017}. \citet{Martin-Navarro2015} measured the 
IMF of a massive relic galaxy, NGC~1277 \citep{Trujillo2014}, and found that it is 
bottom heavy at all radii. \citet{Villaume2017} studied various compact stellar 
systems, and found that despite their large metallicity and [Fe/H] range, they all 
have elevated $\alpha_{\rm IMF}$ with low scatter.  All of these works indicate 
the trend that more compact galaxies are more likely to have bottom heavier IMF.  

What may be the physical explanation?  First of all, simulations \citep[e.g.][]{Chabrier2014} 
reveal that star formation in the extreme environment of very dense and turbulent gas 
will extend the peak mass of the IMF to lower masses.  Compared to massive spiral 
galaxies or less compact massive ETGs, the extreme local environment in the most 
massive compact ETGs could be the reason for their bottom heavy IMF.  Second, 
many simulations and observations support the two-phase formation scenario for 
massive ETGs \citep[e.g.][]{Oser2010,vanDokkum2010, Patel2013}. The idea is that local 
massive ETGs first formed as compact ``red nuggets'' at high redshift due to strong 
dissipational processes.  At later times, they experienced the accretion of 
lower-mass systems.  As a result, they build up their effective radii over time.  
Compactness therefore tracks the relative fraction of the stars formed in the 
first phase \citep[e.g.][]{Bezanson2009}.  The compact galaxies in our 
sample may preserve more of the properties of the 'first phase' formation at high 
redshift and have been less disturbed by the minor mergers at low redshifts.  
%The IMF of the less compact galaxies are affected by the subsequently merged satellite galaxies.  
For the less compact galaxies the stellar IMF is affected by both in-situ 
and ex-situ activity. Our results are consistent with the picture that 
both the formation time and the star formation timescale are related to 
the stellar IMF in the galaxy centers.  
%Considering the relation between [Mg/Fe] and environment, the stellar IMF in 
%those merged systems may carry environmental information 
%as well  

Our next step is to study the local IMF of massive ETGs as a function of radius. 
Previous studies have revealed steep gradients in the metallicity profile with radius, 
and in general a nearly flat [Mg/Fe] radial profile in massive ETGs. 
The latter could be due to the environmental quenching of low-mass galaxies, 
which were accreted and distributed at the outskirts of massive galaxies 
\citep[e.g.][]{Gu2018b,Gu2020}.  Due to the differences in the gradients, 
we expect to see a radial dependence from the relations among metallicity, 
[Mg/Fe], and IMF.  In the next paper we will focus on their relations at different 
fractions of $R_{\rm e}$.  How the correlations hold at different radii will help 
us understand which of the stellar population properties are fundamental, and the
role of in-situ and ex-situ processes in driving IMF variation.

Our current work is limited by the dynamic range in stellar mass of the 
MASSIVE sample.  Although we include low mass galaxies from CvD in our analysis,
the stellar populations studied are restricted to total metallicity and [Mg/Fe].  
\S~4.2 shows the trend that [Na/Fe] and [O/Fe] both increase with increasing 
central velocity dispersion (Figure~\ref{fig_sigmarel}).  
In the future, through studying low mass ETGs with a consistent model we will 
be able to tell whether other elemental abundances are related to IMF variations.

% ---------------------------------------------------------------------------- %
\begin{figure*}[th]
\vskip 0.15cm
  \centerline{\psfig{file=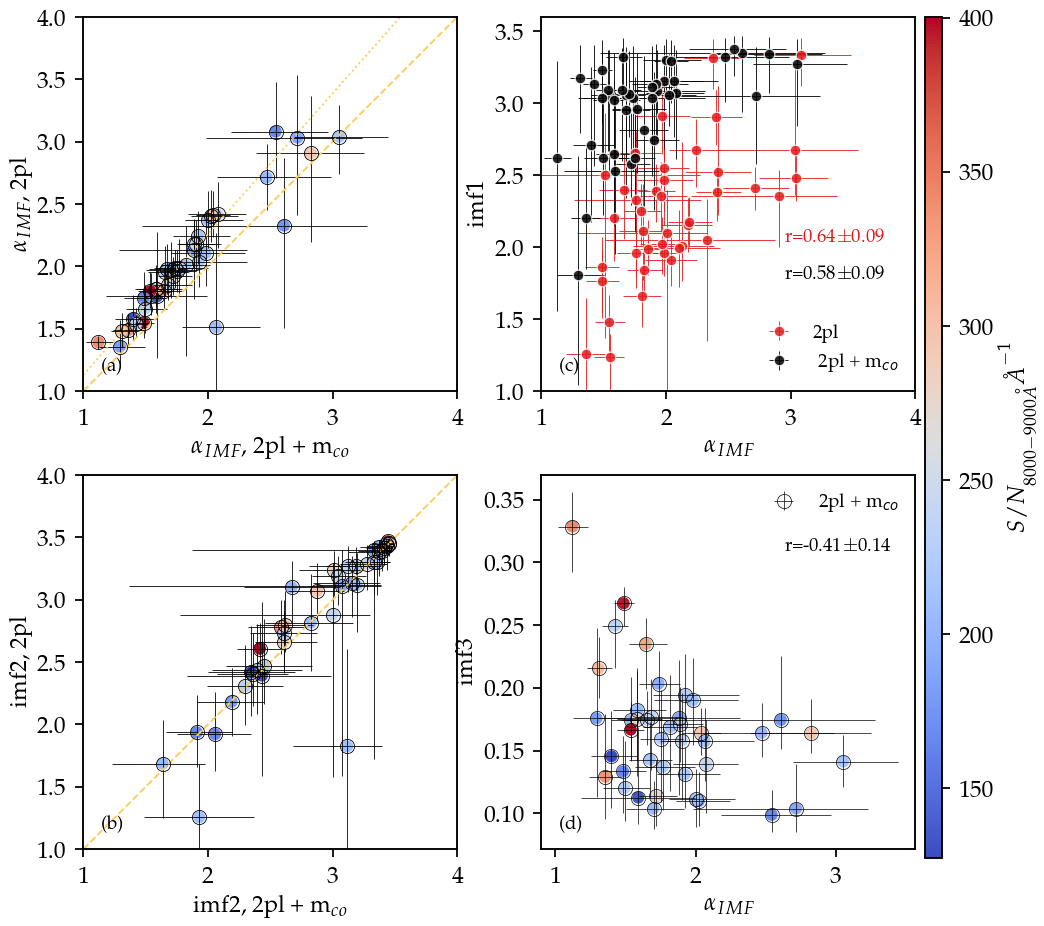, 
  width=15cm}}
\caption{ Comparison of IMF parameters from two assumptions of IMF model: 
(1) double power-law with a fixed low-mass cutoff of $0.08M_{\odot}$, and (2) 
double power-law with variable low-mass cutoff (fiducial model in this work).  
{\tt imf1} and {\tt imf2} are the low and intermediate mass IMF slopes, respectively, 
while {\tt imf3} is the low mass cutoff. (a) and (b): comparison of $\alpha_{\rm IMF}$ 
and {\tt imf2} between the two models: they are overall consistent, but the $M/L_r$ 
in the double power-law model is $12\%$ higher in $M/L_r$ compared to our 
fiducial model (orange dotted line). 
(c) and (d): {\tt imf1} and {\tt imf3} as a function of $\alpha_{\rm IMF}$, 
and the corresponding Pearson correlation coefficients.  
}
\label{fig_modelcomp}
\end{figure*}
\noindent  
% ---------------------------------------------------------------------------- %

\subsection{IMF Model Comparison}

As indicated by \citet{Newman2017}, the parameterizations of IMF models are 
important in spectral modeling, and different functional forms 
(e.g., single or double slopes, with or without a low-mass cutoff) of the
IMF may lead to different inferred $M/L$.   As described in \S~2, our fiducial IMF model 
has a slope above $1M_{\odot}$ fixed to 2.3, and free parameters {\tt imf1} and 
{\tt imf2} for the slopes below and above $0.5M_{\odot}$.  In addition, {\tt imf3} 
is used to describe the cutoff mass at the low mass end.  In this section, 
we test the results of different IMF parameterizations by comparing our results 
to the assumption that the IMF has the form of a double power-law with a fixed cutoff 
mass at $0.08M_{\odot}$, the canonical hydrogen-burning limit.  

We compare results from these two models in Figure~\ref{fig_modelcomp}.  In the left panels, 
we compare the median (error-bars indicate the 16th and 84 percentiles) of the posteriors of 
$\alpha_{\rm IMF}$ (top), and the slope in the intermediate mass range ({\tt imf2}) (bottom).  
The y-axis represents results from the model with fixed cutoff mass, and those on the x-axis 
represent our fiducial model.  In general they are consistent with each other.  
The average ratio of $M/L_r$ and $\alpha_{\rm IMF}$ is $11\%$ higher in the double 
power-law model than our fiducial model with very little scatter.  In general the fixed 
low cutoff mass result in slightly higher $M/L$, indicating that our results do not depend 
heavily on the choice of the IMF functional form.  
Adopting the IMF model with fixed cutoff mass will not alter our 
conclusions about the global trend that the IMF in ETG centers becomes more bottom 
heavy with increasing stellar mass.
As shown in Figure~\ref{fig_modelcomp}(b) the slopes in the intermediate mass 
range are consistent with each other, suggesting that the intermediate mass slope 
is not very sensitive to the IMF form.  

Figure~\ref{fig_modelcomp}(c) shows the low-mass IMF slope, {\tt imf1}, in our 
fiducial model (black) and the model with fixed cutoff mass (red) as a function of 
$\alpha_{\rm IMF}$.  Panel (d) shows cutoff mass in our fiducial model as a function of 
$\alpha_{\rm IMF}$.  There is apparently some degeneracy between the low mass 
slope $<0.5M_{\odot}$ {\tt imf1} and the low mass cutoff {\tt imf3}.  By allowing 
{\tt imf3} to vary, we estimate that $\langle${\tt imf3}$\rangle=0.16\pm0.05M_{\odot}$.  
As indicated by the Pearson correlation coefficient at the lower right corners, 
{\tt imf1} is moderately positively correlated with $\alpha_{\rm IMF}$, and 
{\tt imf3} is negatively correlated with $\alpha_{\rm IMF}$, which makes sense since 
allowing the low cutoff mass to vary will result in decreasing $M/L$ with increasing 
cutoff mass. On the other hand if we fix {\tt imf3} (panel (c), red), 
$\alpha_{\rm IMF}$ primarily depends on {\tt imf1} with a correlation coefficient of $0.64$.  
{\tt imf2} is stable (panel (d)) with and without a flexible low cutoff mass, 
suggesting that a free intermediate mass slope is necessary in the IMF model, 
and a double power-law IMF form is a better choice than a single 
power law.  For high S/N galaxies the estimated low-mass cut-off values, {\tt imf3}, 
are still larger than $0.08M_{\odot}$, suggesting that {\tt imf3} is also a useful free 
parameter in describing the IMF at low mass range.
However, we do not intend to use this as an accurate measurement of the cutoff mass. 
We simply use {\tt imf3} as an additional free parameter, since a fixed $0.08M_{\odot}$ 
does not apply for all galaxies and may over-estimate the $M/L$ and $\alpha_{\rm IMF}$. 
If we switch our results to the $2pl$ model, the positive correlations between 
$\log(\alpha_{\rm IMF})$ and [Mg/Fe], [Z/H], $\log(\sigma)$ are still strongly held, 
although the $\log(\Sigma_{\rm Kroupa})-\log(\alpha_{\rm IMF})$ within our sample becomes 
less significant with $r=0.27\pm0.16$

\section{Summary}

We have conducted detailed full spectral modeling on a sample of 41 massive 
early-type galaxies in the volume-limited MASSIVE survey to constrain their stellar 
populations and the stellar initial mass function.  Galaxies in our sample 
are among the most massive in the universe.  We extract spectra observed 
by LDSS-3 on the Magellan/Clay telescope, using an effective circular aperture 
of $R_{\rm e}/8$.  We obtain high S/N spectra with $\langle S/N\rangle=234$\AA$^{-1}$ in $0.8-0.9\mu m$, 
and fit for stellar population parameters, $M/L$ and IMF mismatch parameter, $\alpha_{\rm IMF}$. Our main results are summarized as follows:

\begin{enumerate}

\item In our default model, the stellar IMF is described by three free parameters: 
the low and intermediate mass slopes, and a low mass cutoff.  Spectral modeling using 
an IMF fixed to Kroupa results in visibly worse residuals and fails to describe 
the spectra in the centers of massive early-type galaxies.

\item Within $R_{\rm e}/8$, the IMF of all galaxies in our sample are more bottom heavy 
than Kroupa.  The $\alpha_{\rm IMF}$ mismatch parameter of the whole sample is 
$\langle \alpha_{\rm IMF}\rangle=\langle(M/L)/(M/L)_{\rm MW}\rangle=1.84\pm0.43$. On average 
these massive galaxies have stellar IMF bottom heavier than the IMF with a 
Salpeter slope. 

\item Combining the results of galaxies in our sample with lower-mass ETGs in the
previous literature, we confirm the positive trend that the central IMF becomes bottom 
heavier with increasing galaxy central velocity dispersion, stellar mass, luminosity 
and dynamical mass (\S~4.4). 
We find correlations between $\log(\alpha_{\rm IMF})$ and, $\log(\sigma)$, 
[Mg/Fe], and total metallicity ([Z/H]) (\S~4.4).  Within our sample, 
$\log(\alpha_{\rm IMF})$ is 
positively correlated with both the effective surface mass density ($\Sigma$) and 
total metallicity (\S~4.3), suggesting that galaxy compactness might be 
an important property related to IMF variation.

\item We estimate the dynamical masses and compare them with the stellar mass 
within $R_{\rm e}$.  Most of the galaxies have stellar mass consistent within 
$1\sigma$ uncertainty with the estimated dynamical mass (\S~5).
Most galaxies have central $M/L$ within $R_{\rm e}/8$ 
higher than the average dynamical $M/L$ within $R_{\rm e}$, and thus choice of 
apertures is important in the comparison.

\item The compact galaxies in our sample may preserve more of the properties from their 
'first phase' of formation.  Our results are consistent with the picture that both the 
formation time and the star formation timescale are related to the stellar IMF in the 
galaxy centers.  

\end{enumerate}

In an upcoming paper, we will look into the radial dependence of the 
IMF and stellar populations, and study the connections between these local properties.  
This is important since we know there are usually strong gradient in metallicity, 
and the radial variation of population properties such as $\alpha$-abundances and stellar 
age might be sensitive to the large scale environment.  In the future, a larger sample 
of lower mass galaxies will help with the investigation of the physical mechanisms driving 
IMF variations over a larger dynamical range, and hopefully will provides us insights 
on any connections between galaxy stellar IMF with properties other than [Mg/Fe] and [Z/H]. 
A large sample of lower mass ETGs including both compact and diffuse ETGs will 
help verify how significant and on what scale galaxy compactness is related to galaxy IMF.

\acknowledgments
This paper includes data gathered with the 6.5 meter Magellan Telescopes located 
at Las Campanas Observatory, Chile.  The authors are pleased to acknowledge that the work reported on in this paper was substantially performed using the Princeton Research Computing resources at Princeton University which is consortium of groups led by the Princeton Institute for Computational Science and Engineering (PICSciE) and Office of Information Technology's Research Computing

This project used data obtained with the Dark Energy Camera (DECam), which was constructed by the Dark Energy Survey (DES) collaboration. Funding for the DES Projects has been provided by the U.S. Department of Energy, the U.S. National Science Foundation, the Ministry of Science and Education of Spain, the Science and Technology Facilities Council of the United Kingdom, the Higher Education Funding Council for England, the National Center for Supercomputing Applications at the University of Illinois at Urbana-Champaign, the Kavli Institute of Cosmological Physics at the University of Chicago, Center for Cosmology and Astro-Particle Physics at the Ohio State University, the Mitchell Institute for Fundamental Physics and Astronomy at Texas A\&M University, Financiadora de Estudos e Projetos, Fundacao Carlos Chagas Filho de Amparo, Financiadora de Estudos e Projetos, Fundacao Carlos Chagas Filho de Amparo a Pesquisa do Estado do Rio de Janeiro, Conselho Nacional de Desenvolvimento Cientifico e Tecnologico and the Ministerio da Ciencia, Tecnologia e Inovacao, the Deutsche Forschungsgemeinschaft and the Collaborating Institutions in the Dark Energy Survey. The Collaborating Institutions are Argonne National Laboratory, the University of California at Santa Cruz, the University of Cambridge, Centro de Investigaciones Energeticas, Medioambientales y Tecnologicas-Madrid, the University of Chicago, University College London, the DES-Brazil Consortium, the University of Edinburgh, the Eidgenossische Technische Hochschule (ETH) Zurich, Fermi National Accelerator Laboratory, the University of Illinois at Urbana-Champaign, the Institut de Ciencies de l'Espai (IEEC/CSIC), the Institut de Fisica d’Altes Energies, Lawrence Berkeley National Laboratory, the Ludwig Maximilians Universitat Munchen and the associated Excellence Cluster Universe, the University of Michigan, NSF's NOIRLab, the University of Nottingham, the Ohio State University, the University of Pennsylvania, the University of Portsmouth, SLAC National Accelerator Laboratory, Stanford University, the University of Sussex, and Texas A\&M University.
% --------------------------------------------------------------------------- %
% this part is for using the legacy survey data

% ---------------------------------------------------------------------------- %
\appendix
\begin{figure*}[hb]
\vskip 0.15cm
  \centerline{\psfig{file=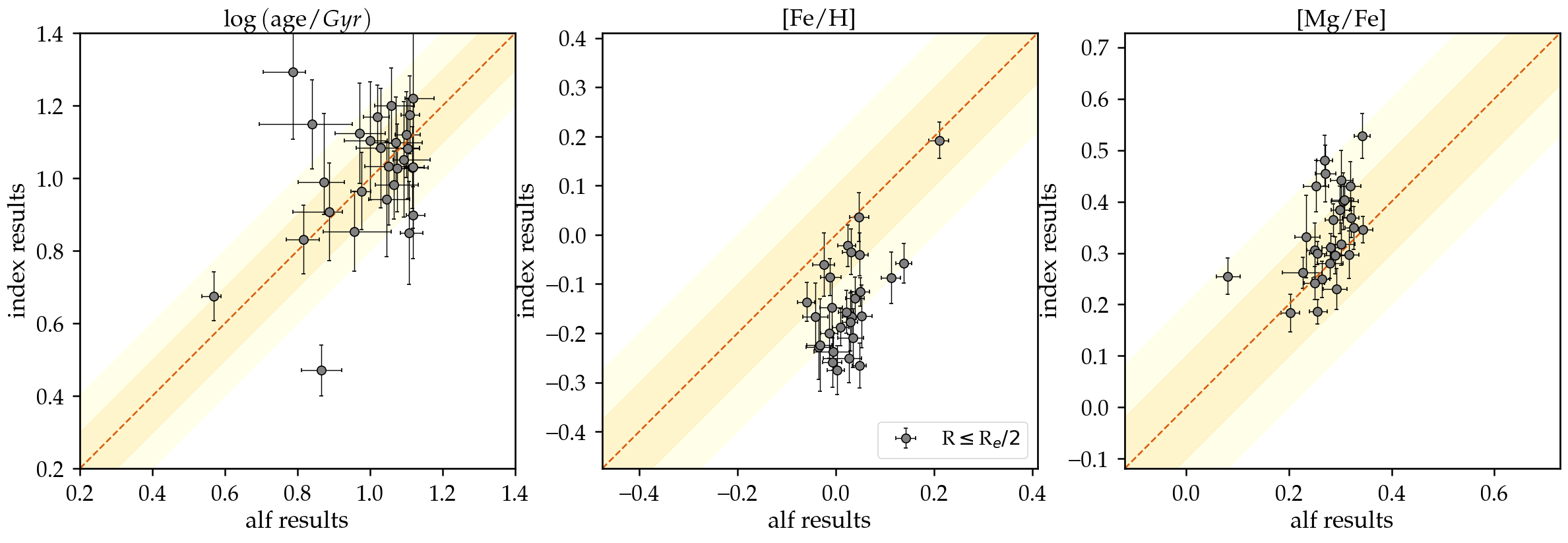, width=18cm}}
  \caption{Comparison between full spectral modeling results from {\tt alf} 
  in this work, and results from Lick indices measured on IFS data.  
  Orange and yellow regions indicate a difference of 0.1~dex and 0.2~dex, 
  respectively.
 }
\end{figure*}
%\include{Appendix_A}
%\section*{Comparing stellar population results with \citet{Greene2019}}
% Figure1: compare with MASSIVE XII
% todo: missing galaxies in MASSIVE XII sample:
% NGC3209, NGC3462, NGC3862, NGC7052, NGC7619, NGC7626

In this section we compare the stellar populations: stellar age, [Fe/H] 
and [Mg/Fe] in 28 ETGs in this work and \citealt{Greene2019} (MASSIVE Paper XII).  
The stellar 
populations in MASSIVE Paper XII are measured with Lick indices on 
spectra observed with the Mitchell IFS at the McDonald Observatory. 
The wavelength ranges in indices measurement is 3650--5850\AA.
In the comparison, we use a consistent aperture of $R_{\rm e}/2$, where $R_{\rm e}$ are 
measured on CFHT $K-$band imaging (Quenneville in prep).  The results 
are shown in Figure~11.  The index results are the mean stellar populations 
within $R_{\rm e}/2$, while the alf results are derived modeling a stacked 
spectra within an effective circular aperture of $R_{\rm e}/2$.  
On average, the offsets are 
$\langle$[Fe/H]$_{\rm index}\rangle - \langle$[Fe/H]$_{\rm alf}\rangle=-0.16$~dex,
$\langle\log({\rm age}/{\rm Gyr})_{\rm index}\rangle - \langle\log({\rm age}/{\rm Gyr})_{\rm alf}\rangle=0.015$~dex and $\langle$[Mg/Fe]$_{\rm index}\rangle - \langle$[Mg/Fe]$_{\rm alf}\rangle=0.06$~dex.  The overall agreement is encouraging.  On average the 
[Fe/H] in our results is higher than the index results from MASSIVE Paper XII. 
We note that in addition to the differences in model and modeling methods, 
there are differences in wavelength ranges, data quality, and spatial information 
in these two works.

% --------------------------------------------------------------------------- %
\vspace{1cm}
\bibliography{references}

\end{document}